\documentclass[10pt, journal, letterpaper]{IEEEtran}
\usepackage{amsmath,amsthm,dsfont}
\usepackage{tabularx,multirow}
\usepackage{graphicx,subfigure,comment}
\usepackage[table]{xcolor}
\usepackage{url} 
\usepackage{dsfont}
\usepackage{cite}
\usepackage{siunitx}
\usepackage{physics} 
\usepackage{pstool}
\usepackage{enumitem}
\usepackage{algorithm}
\usepackage[noend]{algpseudocode}

\newcommand{\Fig}[1]{Fig.~\ref{fig:#1}}
\newcommand{\Prope}[1]{Property~\ref{prope:#1}}
\newcommand{\Propo}[1]{Proposition~\ref{propo:#1}}

\newcommand{\Sec}[1]{Sec.~\ref{sec:#1}}
\newcommand{\Tab}[1]{Tab.~\ref{tab:#1}}
\newcommand{\Eq}[1]{(\ref{eq:#1})}
\newcommand{\Alg}[1]{Alg.~\ref{alg:#1}}
\newcommand{\Line}[1]{Line~\ref{line:#1}}

\newcommand{\ind}[1]{\mathds{1}_{\left[{#1}\right]}}
\newcommand{\Lc}{\mathcal{L}}
\newcommand{\Nc}{\mathcal{N}}
\newcommand{\Dc}{\mathcal{D}}
\newcommand{\Ic}{\mathcal{I}}

\newcommand{\Tc}{\mathcal{T}}
\newcommand{\Tb}{\mathbf{T}}
\newcommand{\Eb}{\mathbf{E}}
\newcommand{\Kb}{\mathbf{K}}

\newcommand{\wb}{\mathbf{w}}
\newcommand{\yb}{\mathbf{y}}

\newtheorem{property}{Property}
\newtheorem{proposition}{Proposition}

\def\Herm{^\mathsf{H}}

\allowdisplaybreaks

\begin{document}

\title{
Efficient Distributed DNNs\\in the  Mobile-edge-cloud Continuum
} 

\author{Francesco~Malandrino,~\IEEEmembership{Senior~Member,~IEEE,}
Carla~Fabiana~Chiasserini,~\IEEEmembership{Fellow,~IEEE,}
Giuseppe~Di~Giacomo
\thanks{F.~Malandrino and C.~F.~Chiasserini are with CNR-IEIIT and CNIT, Italy. C.~F.~Chiasserini and G.~Di~Giacomo are with Politecnico di Torino, Italy.}
\thanks{A preliminary version of this work was presented in~\cite{noi-wons22}.\\
This work was partially Funded by the NPRP-S 13th Cycle grant no.\,NPRP13S-0205-200265 from the 
Qatar National Research Fund (a member of Qatar Foundation), 
and by European Union under 
grant agreement no.\,101069688. 
Views and opinions expressed are however those of the authors only and do not necessarily 
reflect those of the European Union. Neither the European Union nor the granting authority 
can be held responsible for them.}
} 
\maketitle

\begin{abstract}
In the mobile-edge-cloud continuum, a plethora of heterogeneous data
sources and computation-capable 
nodes are available. Such nodes can cooperate to perform a {\em
distributed} learning task, aided by a learning controller (often
located at the network edge). The controller is required to make decisions
concerning (i) data selection, i.e., which data sources to use; (ii)
model selection, i.e., which machine learning model to adopt, and (iii)
matching between the layers of the model and the available physical
nodes. All these decisions influence each other, to a significant extent
and often in counter-intuitive ways. In this paper, we formulate a problem addressing all of the above aspects and 
present a
solution concept called RightTrain, aiming at making the aforementioned
decisions in a joint manner, minimizing energy consumption subject to
learning quality and latency constraints. RightTrain leverages an expanded-graph representation of the system and 
a delay-aware Steiner tree to obtain a provably near-optimal  solution while keeping the time complexity low. 
Specifically, it runs in polynomial
time and its decisions exhibit a competitive ratio of $2(1+\epsilon)$,
outperforming state-of-the-art solutions by over 50\%.
Our approach is also validated through a real-world implementation. 
\end{abstract}


\section{Introduction}
\label{sec:intro}

Enabling technologies like 5G networks and distributed machine learning (ML) have fostered the emergence of the so-called {\em Internet of Intelligent Things} networking paradigm, allowing user equipment (UEs), e.g., smartphones or smart-city actuators, to leverage cloud-based artificial intelligence services.
This scenario is expected to further evolve towards the 
edge intelligence paradigm~\cite{peltonen20206g,wang-netmg-19}: ML-based applications
will move from remote, cloud-based servers to the mobile network,
including computation-capable devices at the network edge and mobile
UEs. Indeed, recent reports~\cite{baresi2019unified} highlight how the capability of edge and
mobile devices is growing much faster than cloud ones, soon leading to 
complete interoperability among mobile, edge, and cloud and, thus, to the formation of a {\em
continuum}.

The ML tasks to perform will be as diverse as the devices performing them. Indeed, the best ML approach to adopt depends upon such factors as the application and scenario at hand, as well as the type and quantity of available data. Possible options include 
supervised~\cite{caron2018deep,kappeler2015combining,zhuang2019local}
and  unsupervised learning (most notably, deep domain adaptation
(DDA)~\cite{mancini2019inferring,loghmani2020unsupervised,
planamente2021da4event}), as well as hybrid applications
combining labeled and unlabeled data through multiple learning and pseudo-labeling 
techniques~\cite{albaseer2020exploiting}.
Many of these
tasks can be accomplished through deep neural networks (DNNs), built by
combining a sequence of {\em layers} of different types. 
The DNNs used for unsupervised and hybrid learning tend to be more complex than their
supervised learning counterparts, however, they use the very same
building blocks, e.g., convolutional and fully-connected
layers~\cite{caron2018deep,zhuang2019local}. 
The possibility of using a relatively small set of building blocks to
support a wide set of ML-based applications has motivated the ML-as-a-Service
(MLaaS), 
whereby the network provides a
set of {\em customized} ML-based services, e.g., image recognition or
clustering, to the applications using the network resources.  

Owing to the complexity of the learning tasks to perform, as well as to
the need to keep as much as possible the information coming from different sources local, 
it is expected that most of
MLaaS learning will be performed in a {\em distributed} fashion.
Distributed learning is an excellent fit for edge intelligence
scenarios, as it envisions leveraging the data and resources of multiple
nodes in order to perform a common learning task. 
Additionally, recent works on ML techniques tailored for lower-powered
devices~\cite{lane2016deepx,cai2020tinytl} has significantly extended
the types of devices that can be leveraged for distributed ML, along
with their diversity. 
We thus argue that ML-based services
can seamlessly  use any of the nodes of the
mobile-edge-cloud continuum, so as to ensure that requirements (e.g.,
learning time) are met.

Among the factors influencing the performance of MLaaS, combined with distributed learning,
the most prominent are (i) the quantity and diversity of the input data used for training; (ii) the actual learning strategy employed, e.g., the layers composing the DNN; (iii) the resources, (e.g., computational) available to the learning task. 
Such
decisions strongly impact one another, e.g., using more data requires
more resources to promptly process them; thus, it is important that
they are made in a  joint manner. The split learning (SL)
paradigm~\cite{vepakomma2018split,gupta2018distributed,gao2020end},
envisioning to run a subset of the DNN layers at each level of the
network topology, represents a significant step in this direction. However, SL
is concerned only with placement decisions
and, since it only splits DNN into as many parts as there are network topology layers 
(e.g., one for edge and one for cloud), may not always be able to 
reap the full benefits of the mobile-edge-cloud continuum.

Another limit of state-of-the-art works on distributed ML 
is their emphasis on learning {\em effectiveness}, e.g., classification
accuracy, over {\em efficiency}. Indeed, most existing approaches are
concerned with locating and selecting the highest possible quantity of
computational resources, and use them to process the largest possible
quantity of data, in order to obtain the highest-quality possible
learning. However, awareness is rising that 
seeking the utmost performance is not
necessarily the most desirable strategy in real-world situations, and the
concept of ML  efficiency -- as opposed to sheer performance -- is
rapidly gaining prominence~\cite{green-ai}.
In fact, while {\em inference} has a relatively minor impact on 
the total resource consumption, the {\em training} of a DNN is very demanding 
in terms of processing power, hence, energy.

In this paper, we address the above issues by presenting RightTrain, a decision-making framework allowing joint, high-quality decisions on (i) which data to use for learning, (ii) which DNN structure to employ, and (iii) which physical nodes and resources therein to use. 
Our framework can capture the nontrivial (and, often,
counter-intuitive) ways in which such choices interact with each other,
and yields decisions that are provably close to the optimum, while
keeping a low computational complexity.

More specifically, our contributions can be summarized as follows:
\begin{itemize}
    \item we propose  a model that describes the components of
a MLaaS system based on DNNs, their behavior, and their inter-dependencies;

    \item we formulate the problem of {\em jointly} making decisions on:
      (i) 
the data to be used for model training, (ii) the structure
of the DNN to adopt, and the resources to allocate for the learning process, 
with the aim of minimizing the energy consumption 
while meeting a target maximum learning time and desired learning quality;

    \item we present a solution, called {\em RightTrain}, solving
the above problem in polynomial time and yielding provably near-optimal
(namely, with a $2(1+\epsilon)$~competitive ratio) solutions;

    \item we compare RightTrain against the SL state-of-the-art approach,
and find how the greater flexibility of RightTrain results in a
significantly lower energy consumption, especially when the target
learning time is not too short;

\item we show the feasibility of our approach using a lab test-bed implementation.
\end{itemize}
As discussed in \Sec{relwork}, several existing works have tackled the
problem of selecting the computational and network resources needed for
a given learning task, and a few have studied the impact of
different DNN structures on the learning performance. Our work
is, however, the first to identify and solve the important challenge of
{\em adapting} the DNN structure, the network resources and the size of the datasets to
one another, thus achieving unparalleled learning 
efficiency and performance.

The remainder of this paper is organized as follows. After discussing
related work in \Sec{relwork} 
and summarizing our main results in \Sec{roadmap}, 
we describe our system model and problem
formulation in \Sec{model} and our characterization of learning performance
in \Sec{characterizing}. We then introduce the RightTrain solution concept in
\Sec{algo}, and formally prove its complexity and competitive ratio
properties in \Sec{analysis}. Finally, 
\Sec{peva} shows the
performance of RightTrain against the state-of-the-art,
\Sec{testbed} reports how we validate
our model and approach through a lab test-bed,
and \Sec{conclusion} concludes the paper.

\section{Related Work}
\label{sec:relwork}

A first body of works related to ours~\cite{neglia,wang2019adaptive,merluzzi2020dynamic} 
target the problem of characterizing the performance of distributed ML, 
accounting for such aspects as the topology of the {\em logical} network formed by
cooperating nodes~\cite{neglia} and the
computational~\cite{wang2019adaptive,merluzzi2020dynamic} and
communication~\cite{merluzzi2020dynamic} resources they are assigned.

Concerning distributed learning techniques themselves, one of the most
prominent is federated learning (FL)~\cite{konen2015federatedOptimization}, 
whereby participating nodes train the same DNN with their local data, and
send the resulting weights to a coordination server that averages the weights 
and sends them back to the nodes.  FL has become one of the most popular approaches
to cooperative learning also in mobile~\cite{kang2020reliable} and
MEC-based~\cite{wang2019adaptive} scenarios. 
Since nodes are expected to contribute their own resources to
learning, incentive mechanisms may be necessary to foster
cooperation~\cite{zhan2020learning},
also employing blockchain~\cite{kim2019blockchained}.

Several recent works have endeavored to characterize and improve the performance of FL under realistic conditions, most notably, learning nodes with heterogeneous datasets and/or capabilities. The authors of~\cite{li2019convergence} consider the classic FedAvg strategy, and characterize its performance, e.g., the loss reduction as a function of the number of local epochs and global iterations.
The later work~\cite{nguyen2020fast} aims at going beyond FedAvg and proposes an alternative strategy called FedUN. FedUN optimizes the convergence speed by choosing (``sampling'') the learning nodes to use at each epoch -- and weighting their updates -- based upon local gradient information. \cite{nguyen2020fast}~also provides a lower bound for FedUN's loss reduction at each epoch, hence, the total convergence time.
\cite{li2019fair}~takes a different approach and tackles the issue of fairness, i.e., how to ensure that devices with different capabilities and/or datasets have similar learning performance. The strategy envisioned in~\cite{li2019fair} is predicated upon giving {\em more} weight to updates from the nodes with the worst performance, resulting in a better learning quality for such nodes at the cost of a higher number of global iterations.

Split learning (SL) is a recently-emerging paradigm predicated on
partitioning the DNN among the nodes participating in the
learning process. SL drops FL's requirement that all nodes have the same
DNN, and has been found to outperform ML in a wide variety of
scenarios~\cite{gao2020end}, owing to its ability to match the learning
operations and the hardware performing them. Other works envision
similar approaches, based on choosing the right network node to run each
layer of a DNN and accounting for device capability~\cite{zhou2019edge},
network latency~\cite{zhou2019edge,li2019edge}, and
privacy~\cite{zhou2019edge,mao2018privacy}. 
\cite{li2019edge}~takes a further step, combining DNN splitting with ``right-sizing'', i.e., 
removing some DNN layers if they are not necessary to reach the required learning quality. 
However, that work only focuses on inference, and neither the quantity of data to use 
nor the resources to assign are accounted for.

Among the few works jointly making learning- and network-related
decisions, \cite{zeng2019boomerang}~aims at (i) right-sizing the DNN for
the task at hand, i.e., skipping some layers if the classification
precision is sufficiently high, and (ii) offloading a part of the DNN to
edge-based nodes. Notice, however, that \cite{zeng2019boomerang}~only
supports DNNs with a chain topology, and does not support the use of multiple
sources of information.
Still in the context of inference, the authors of~\cite{mohammed2020distributed} envision partitioning the DNN into an arbitrary number of parts, and running each of them at the most appropriate node in the edge-cloud continuum.
In a similar setting,
\cite{baccour2020distprivacy} addresses security issues, 
placing different layers of an image-classification DNN on different
devices, preventing any device from seeing enough layers to
reconstruct the original image.

Recent work~\cite{salem2021towards} targets a heterogeneous scenario similar to the mobile-edge-cloud continuum, and envisions an {\em inference delivery network} whereby each node offers one or more ML models. Inference tasks -- which are assumed to require one of several alternative models, e.g., DNNs with different architectures -- can then be carried out at the most appropriate node, accounting for both cost and delay issues. Compared to~\cite{salem2021towards}, our work (i) targets the {\em learning} phase, which is the most challenging and resource-intensive; (ii) allows breaking down {\em one} model (e.g., one DNN) across multiple devices, and (iii) does not depend upon the assumption that alternative models exist for a given task.

Our work is also related to the recent but growing body of works
accounting for ML energy consumption, and the
resulting carbon footprint. As reported in \cite{assran19},
training one complex  ML model may lead to a carbon footprint
equivalent to 5 times the lifetime emissions of an average
car. Thus, it is critical to envision solutions that, exploiting
the edge intelligence concept \cite{wang-netmg-19,zomaya20},
effectively exploit the physical proximity of a large number of devices,
each collecting data and equipped with computational and memory
resources. At a higher level, as advocated in~\cite{green-ai}, this
calls for a different view of ML goals where the focus shifts
from sheer learning quality (e.g., classification accuracy) to the more
comprehensive concept of {\em efficiency}.

Finally, a preliminary version of RightTrain has been presented in our conference paper~\cite{noi-wons22}. 
Compared to~\cite{noi-wons22}, this paper includes a more rigorous discussion of RightTrain and the principles it is based upon, a formal characterization of its computational complexity and competitive ratio, additional performance evaluation scenarios, and a testbed-based validation.

{\bf Novelty.}
Our holistic approach contributes to making ubiquitous ML reality, {\em jointly} 
addressing
issues that earlier have  been only marginally or incidentally dealt with. 
Specifically, we account for the mutual influence of the
decisions on (i) the data used for DNNs training, (ii) the DNN structure
employed, and (iii) the physical nodes running the latter. Accounting
for all aspects and making all decisions jointly allows us to reach a
level of effectiveness {\em and} efficiency that cannot be matched by
existing approaches.

\section{Main Results and Roadmap}
\label{sec:roadmap}

Our first major contribution is represented by the system model and
problem formulation  summarized in
\Fig{fresco} and described in \Sec{model}. The system model accounts for all the main entities
and aspects involved in distributed training of DNNs over the mobile-edge-cloud continuum. 
It  can describe ML approaches leveraging data parallelism,
model parallelism, or a combination of both -- including split
learning~\cite{vepakomma2018split,gupta2018distributed}.
The problem formulation then formalizes how the decisions of (i) selecting the
input data to be leveraged for learning, (ii) choosing the DNN structure to
be used, and (iii) matching DNN layers with physical servers influence each
other, and determine the learning efficiency under the constraints imposed by both the 
learning process and the network system. The problem is shown to be NP-hard.

Next, through a combination of existing measurements taken from the
literature and our own experiments,  we assess how such decisions impact the
learning performance, namely, learning time, learning quality, and energy consumption
(\Sec{characterizing}). Importantly, so doing, we bridge the gap
between the abstract system model of \Sec{model} and actual, real-world
distributed ML solutions.

Building upon the above results and in light of the problem complexity, 
we envision a solution concept, also
summarized in \Fig{flowchart}. Its main component is the RightTrain
algorithm, detailed in \Alg{mapping} (\Sec{algo}), which leverages expanded graphs
such as the one shown in \Fig{bigtree} and applies a delay-aware Steiner tree
on such graph, to make near-optimal decisions on
data selection, DNN structure, and layer-to-node matching. More
specifically, as proven in \Sec{analysis}, our solution strategy has
polynomial worst-case time complexity (\Prope{complexity}) and a
competitive ratio of~$2(1+\epsilon)$. 
Finally, our performance evaluation shows that the proposed solution reduces by 50\% 
the energy consumption of a learning task when compared to the state of the art, while our lab test-bed implementation shows its feasibility.

\section{System Model and Problem Formulation}
\label{sec:model}

\begin{figure*}
\centering
\subfigure[\label{fig:fresco1}]{
    \includegraphics[width=.2\textwidth]{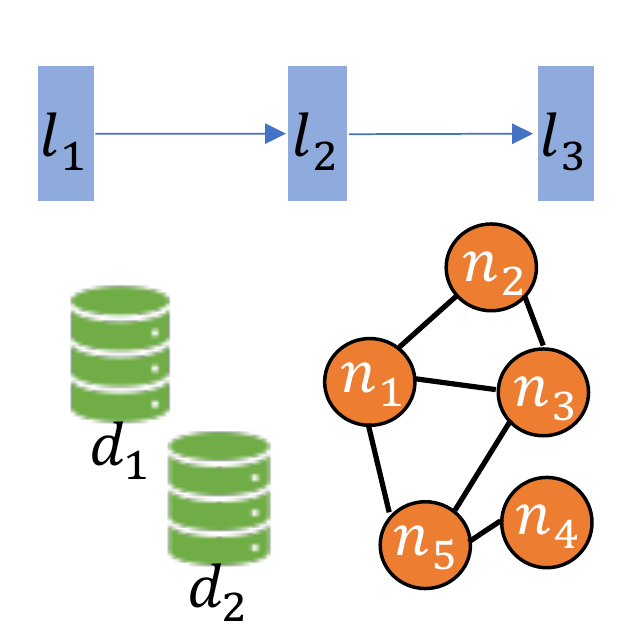}
} 
\hspace{3mm}
\subfigure[\label{fig:fresco2}]{
    \includegraphics[width=.2\textwidth]{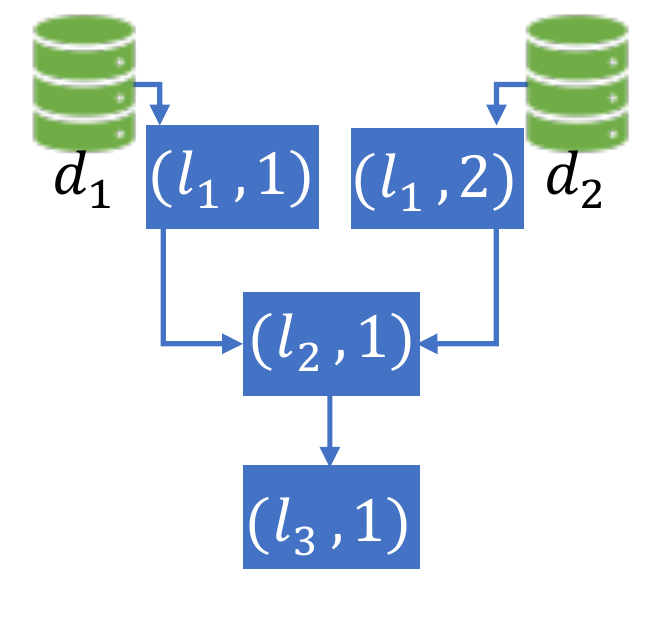}
} 
\hspace{3mm}
\subfigure[\label{fig:fresco3}]{
    \includegraphics[width=.2\textwidth]{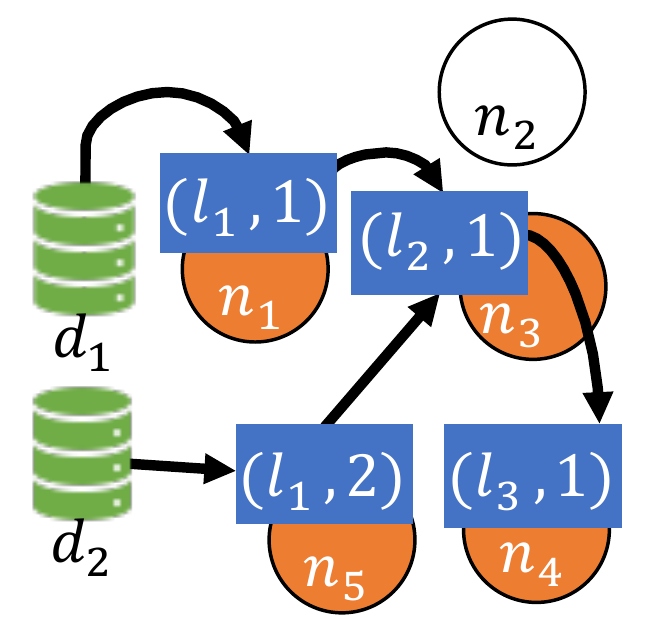}
} 
\hspace{3mm}
\subfigure[\label{fig:fresco4}]{
    \includegraphics[width=.2\textwidth]{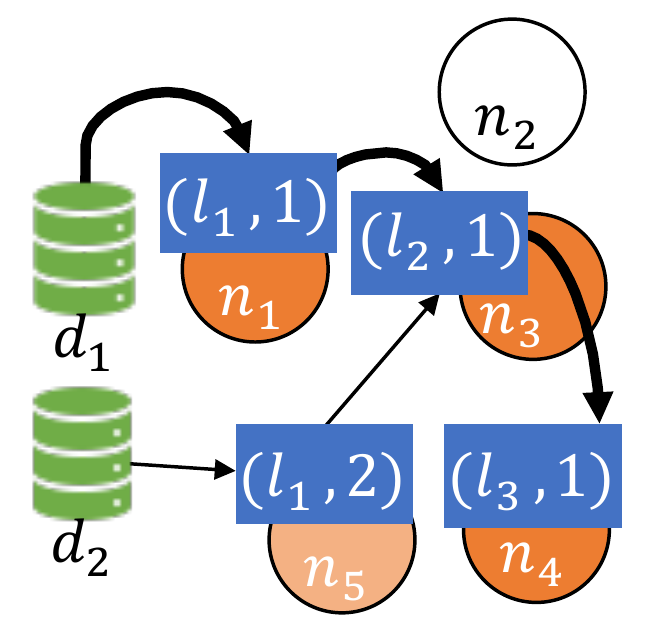}
} 
\caption{
The different stages of the RightTrain approach. (a): input data, namely, a set of DNN {\em layers} (light blue), a set of {\em data sources} (green), and a {\em physical graph} composed of physical nodes (orange). (b): an instance tree, whose nodes are data sources and {\em layer instances}. (c): a possible {\em deployment}, associating layer instances to three out of the five physical nodes. (d): a {\em refined} deployment, using only some of $d_2$'s data and, accordingly, reducing the computational power allocated to layer instance~$(l_1,2)$.
    \label{fig:fresco}
} 
\end{figure*}

Our system model describes a DNN training task leveraging distributed learning, 
and exploiting the resources of multiple mobile, edge, and cloud nodes 
(hereinafter also referred to as learning nodes). 
Such nodes are coordinated by a {\em central
controller}, typically running at the edge of the network infrastructure, 
which can communicate with all learning nodes and  collects information 
on their capabilities and position. 
 The entities the model represents, along with the decisions the central controller has to make, are
depicted in \Fig{fresco}.

\subsection{Input information}

Under the DNN paradigm, learning tasks are performed by a set of {\em
layers} of different types (e.g., fully-connected or convolutional),
organized as a {\em tree}: the learning
result is the output of the tree root, while leaves correspond to data
sources. Each layer has a local set of {\em parameters} that define its
behavior, e.g., the weights of a fully-connected layer, and {\em training} a
DNN means finding the parameter values that minimize a global error
function. As an example, for a classification task, the learning
output~$\yb$ represents the probabilities associated with each class, and
a typically-used loss function is cross-entropy, defined as
$f(\hat{\yb},\yb)=\hat{\yb}^{\Herm} \log\hat{\yb}$, where $\hat{\yb}$ is
a column vector containing the ground truth, and $\Herm$ represents the
transpose operator. Training proceeds in an iterative fashion~\cite{bottou2018optimization} through
several {\em epochs}, each including (i) a {\em forward pass}, where the
input data traverses all instances from the leaves to the tree root, and
(ii) a {\em backward pass} where gradients follow the opposite route
and local parameters at each layer are adjusted so as to reduce the
global loss function. 
The most commonly-employed optimization algorithm is stochastic gradient descent (SGD), though alternatives tailored to ML have been proposed as well~\cite{shamir2014communication}.

Thus, the input to our problem includes (see \Fig{fresco1}):
\begin{itemize}
    \item a set~$\Lc=[l_1,\dots,l_L]$ of {\em DNN layers} 
    connected to
each other to form the DNN structure to implement;
    \item a set~$\Nc$ of {\em physical nodes}, i.e., mobile, edge or
cloud~\cite{baresi2019unified} nodes 
 with the computational capability
to run one (or more) layer instances;
    \item a set~$\Dc$ of {\em data sources}, which may be colocated with physical nodes.
\end{itemize}

For each layer~$l\in\Lc$, we know the computational requirement~$r(l)$,
expressing the amount of computing resources required to process
one unit of traffic entering an instance of~$l$ (e.g., in CPU cycles per
megabit). We are also given coefficients~$q(l)$, denoting the ratio
between outgoing and incoming data for layer~$l$. 

For each node~$n\in\Nc$, we are given the total amount~$R(n)$ of
available computational\footnote{Note 
that a limit on the available energy resources can be included in a
straightforward manner.}
resources therein, that can be shared among all layer instances running
at~$n$. Parameters~$\mu(l,n)\in\{0,1\}$ express whether node~$n$ has
enough memory\footnote{We take memory as representative of non-computing resources; 
any other type of 
resource can be modeled in a similar way.} to execute an
instance of layer~$l$. Concerning data transmission, for each two
nodes~$n_1,n_2\in\Nc$, $S(n_1,n_2)$ indicates the amount of data that
can be transferred over the link between them in a time unit,
with~$S(n_1,n_2)=0$ denoting nodes out of each other's  radio range. 
Finally, let $\Delta(d)$ be the data generated by source~$d\in\Dc$ 
at each epoch, and 
$\eta(d)$ be the node with which $d$ is colocated.

\subsection{Decision variables}
The main decisions to make concern 
(i) how many layer instances to create and how to connect them,
as exemplified in \Fig{fresco2}, (ii) how to deploy the instances onto
the physical nodes, as shown in \Fig{fresco3}, and (iii) how to assign
the computational and network resources, as per \Fig{fresco4}.

{\bf Layer instances and instance trees.} 
For each layer~$l\in\Lc$, we shall create at least one and at
most~$\alpha|\Dc|$ {\em layer instances}, with $\alpha\geq 1$~being a
redundancy factor. Each layer instance runs at a physical node, and it 
is identified as a pair~$(l,i)$, where~$i$ is an index ranging from~$1$
to~$\alpha|\Dc|$, while the set of instances is denoted by $\Ic$. As shown in \Fig{fresco2}, layer instances 
and data sources in~$\Dc$ 
are connected to form an {\em instance tree}, with binary
variables~$y(l,i,m,j)\in\{0,1\}$ expressing whether layer
instance~$(l,i)$ shall be connected to layer instance~$(m,j)$. 
As shown in \Fig{fresco2}, data sources~$d\in\Dc$ are part of the
instance tree, however, each can only be associated with at most one
instance~$(d,1)$. Associating zero instances with a certain data source
means not using it, e.g., because a sufficient quantity of data is
already available.

{\bf Deployment and physical graph.}
Given the  set of layer instances and that of physical 
nodes, we have to decide whether instance~$(l,i)\in\Ic$ shall be deployed 
at node~$n\in\Nc$; such a decision is expressed through binary 
variables~$z(l,i,n)\in\{0,1\}$. We will also identify as~$\nu(l,i)$ 
the node at which instance~$(l,i)$ is deployed, i.e., such that~$y(l,i,\nu(l,i))=1$. 
As for data sources~$d\in\Dc$, 
values $\nu(d,i)$ identify the physical node data source~$d$ is located at.

A further decision  concerns the computational 
resources~$\rho(l,i)\leq R(\nu(l,i))$ to be assigned to each instance~$(l,i)$
and expressed in CPU cycles per second. 
Finally, for each data source $d$,  we have to decide the 
quantity~$x(d,1,m,j)\leq S(\nu(d,1),\nu(m,j))$ of data to be transferred toward layer 
instance $(m,j)$.
We also indicate with~$\chi(l,i,m,j)$ the quantity of data flowing through a 
generic link from instance (or data source) $(l,i)$ to instance~$(m,j)$, defined as:
\begin{equation}
\label{eq:chi}
\chi(l,i,m,j)=\begin{cases}
x(l,1,m,j) & \text{if }l{\in}\Dc \wedge i{=}1,
\\
q(l)\sum_{(h,k)}\chi(h,k,l,i) & \text{otherwise}. 
\end{cases}
\end{equation}

\subsection{Constraints}

The decision variables~$y(l,i,m,j)$, $z(l,i,n)$, $\rho(l,i)$, and
$x(d,1,l,i)$ 
are subject to several constraints. 
Two of them  concern the instance tree exemplified in \Fig{fresco2}
and expressed by the $y$-variables. Specifically, we must deploy at
least one instance of each layer, i.e., 
\begin{equation}
\label{eq:constr-oneinstance}
\sum_{i\in [1\dots \alpha|\Dc|]} \sum_{(m,j)\in\Ic}y(l,i,m,j)\geq
1,\quad\forall l\in\Lc.
\end{equation}
Also, we can only connect on the instance tree {\em subsequent} layers,
i.e., 
\begin{equation}
\label{eq:constr-subsequent}
y(l,i,m,j)\leq \ind{l\text{ is child of }m},
\end{equation}
Then, 
moving to the deployment decisions exemplified in \Fig{fresco3} and
expressed by the~$z$-variables, we must ensure that each instance is
deployed at exactly one physical node:
\begin{equation}
\label{eq:constr-deplonce}
\sum_{n\in\Nc}z(l,i,n)=1,\quad\forall (l,i)\in\Ic.
\end{equation} 
Last, no layer instance can be deployed at a node lacking 
the required memory resources:
\begin{equation}
\label{eq:constr-nocomp}
z(l,i,n)\leq \mu(l,n),\quad\forall (l,i)\in\Ic,n\in\Nc.
\end{equation}

As for the computational resource allocation and data exchange
decisions exemplified in \Fig{fresco3} and expressed by $\rho$ and
$x$-variables, we must ensure that the total amount of resources allocated to all
the instances running at each node~$n$ does not exceed the available
one, i.e.,
\begin{equation}
\label{eq:constr-totpower}
\sum_{(l,i)\in\Ic\colon \nu(l,i)=n}\rho(l,i) \leq R(n),\quad\forall
n\in\Nc.
\end{equation} 
\begin{equation}
\label{eq:constr-totthrp}
\sum_{\substack{(l,i)\colon \nu(l,i)=n \\ (m,j)\colon \nu(m,j)=n'}}
\hspace{-9mm}\chi(l,i,m,j)\leq y(l,i,m,j) S(n,n')
\quad\forall n,n'\in\Nc.
\end{equation}

Finally, we enforce generalized flow conservation~\cite{martin2020okpi},
i.e., the quantity of data going out of layer instance~$(l,i)$ cannot
exceed the product between the quantity of incoming data and~$q(l)$:
\begin{equation}
\label{eq:constr-flow}
\sum_{(m,j)\in\Ic}\hspace{-2mm}\chi(l,i,m,j)\leq
q(l)\sum_{(g,h)\in\Ic}\hspace{-2mm}\chi(g,h,l,i)\quad\forall (l,i)\in\Ic.
\end{equation}
For data sources, the total quantity of outgoing data cannot
exceed~$\Delta(d)$:
\begin{equation}
\label{eq:constr-data}
\sum_{(l,i)\in\Ic} x(d,1,l,i)\leq \Delta(d),\quad\forall d\in\Dc.
\end{equation}

\subsection{Objective function}

Decisions~$x$, $y$, $z$ and $\rho$ fully describe the behavior of
the distributed learning application. However, they do not directly
express: (i) the time taken by each learning epoch, 
(ii)  the energy consumed by each learning epoch, and 
(iii) the number of epochs needed to attain the required loss
function value~$\epsilon^{\max}$. 
We account for these quantities through functions~$\Tb(x,y,z,\rho)$,
$\Eb(x,y,z,\rho)$, and~$\Kb(y,\epsilon)$, respectively. 
The first two are described in \Sec{timecost}, while the third one in
\Sec{accuracy}.

Given functions~$\Tb$, $\Eb$, and~$\Kb$, we formulate our
problem as minimizing the learning energy consumption, subject to
\Eq{constr-oneinstance}--\Eq{constr-data} and to 
achieving the required loss~$\epsilon^{\max}$  by
time~$T^{\max}$:
\begin{equation}
\label{eq:obj}
\min_{x,y,z,\rho}\Kb(y,\epsilon^{\max})\Eb(x,y,z,\rho)
\end{equation}
\begin{equation}
\label{eq:st}
\text{s.t.}\,\,\,\Eq{constr-oneinstance}-\Eq{constr-data}, \,\,\,\Kb(y,\epsilon^{\max})\Tb(x,y,z,\rho)\leq T^{\max}.
\end{equation} 
As we will formally prove in \Sec{analysis}, the above problem is
 NP-hard. 

\section{Characterizing the Learning Performance}
\label{sec:characterizing}

We now show how the performance of the learning process can
be characterized, with reference to the time and energy it takes to perform one epoch
(\Sec{timecost}), and the number of epochs necessary to achieve the required learning quality (\Sec{accuracy}).

\subsection{Epoch duration and energy consumption}
\label{sec:timecost}

DNN layers perform linear algebra operations, hence, it is relatively
straightforward to 
characterize the time they take to perform each epoch, and the
associated energy consumption.
Let us start from epoch time~$\Tb(x,y,z,\rho)$, which depends in a
non-trivial manner upon 
the topology of the instance tree at hand. To compute~$\Tb$, we
first define the computational 
time,~$t_\text{comp}^{l,i}$, taken by layer instance~$(l,i)\in\Ic$:
\begin{equation}
\label{eq:time-one-comp}
t_\text{comp}^{l,i}=\frac{r(l)}{\rho(l,i)}\sum_{(k,h)\in\Ic}\hspace{-
2mm}\chi(k,h,l,i)\,.
\end{equation}
As per \Eq{time-one-comp}, the computation time is given by the ratio
between the number of operations to perform (e.g., given by the amount of data
to process times the requirement $r(l)$) 
and the quantity $\rho(l,i)$ of computing resources assigned to that instance. 
Note that, by constraint \Eq{constr-totthrp}, the sum in \Eq{time-one-comp} accounts for 
all the data transferred from the children instances $k$ to the parent
instance $l$.
We also define the network time,~$t_\text{net}^{n,n'}$, needed to
transfer data from node~$n$ 
to node~$n'$, which depends upon the quantity of data 
transferred from the children instances $k$ running at $n$ to the parent instance
$l$ running at $n'$: 
\begin{equation}
\label{eq:time-one-net}
t_{\text{net}}^{n,n'} = \frac{ \sum_{(h,k),(l,i)\in\Lc\colon
\nu(h,k)=n,\nu(l,i)=n}\chi(h,k,l,i)}{S(n,n')}\,.
\end{equation}

Then, 
we can compute  
times~$t_\text{begin}^{i,j}$ and~$t_\text{end}^{i,j}$ at which
instance~$(l,i)$ 
starts and ends its computation. Specifically, each instance can only
start its processing when all the data it needs from preceding nodes has
arrived~\cite{neglia}, while its end time is given by
the sum between the layer instance begin time and computing time:
\begin{equation}
\label{eq:t-begin}
t_\text{begin}^{l,i}=\max_{(h,k)\in\Ic}\left[y(h,k,l,i)\left(t_\text{end
}^{h,k}+t_\text{net}^{\nu(h,k),\nu(l,i)}\right)\right]\,, 
\end{equation}
\begin{equation}
\label{eq:t-end}
t_\text{end}^{l,i}=t_\text{begin}^{l,i}+t_\text{comp}^{l,i}\,.
\end{equation}
Finally, the epoch duration,~$\Tb(x,y,z,\rho)$, is given by the end
time of the slowest instance of the last layer, i.e.,
\begin{equation}
\label{eq:t-tot}
\Tb(x,y,z,\rho)=\max_{(l,i)\in\Ic\colon l=L}t_\text{end}^{l,i}\,.
\end{equation}

Estimating the energy consumption associated with ML computational tasks 
has been the focus of a significant body of
research~\cite{garcia2019estimation}. In general, the energy
consumption associated with a task (in our case, a learning instance
running at a node) is determined by its usage of CPU, memory, and GPU
resources~\cite[Eq.~(1)]{henderson2020towards}. Those, in turn, depend
upon the layer implemented,  the quantity of data it processes,
and  the characteristics of the node itself. Thus, 
for each layer instance~$(l,i)\in\Ic$, we can write:
\begin{equation}
\label{eq:energy-v}
E^{l,i}_\text{comp}=t_\text{comp}^{l,i}\left[e_\text{p}(\nu(l,i))\rho(l,
i){+}e_\text{f}(l,\nu(l,i))\right].
\end{equation}
In \Eq{energy-v}, we recall that $t_\text{comp}^{l,i}$ is the computation
time for each layer instance at each epoch. Such a time is multiplied by
the power consumed by the node~$\nu(l,i)$ at which the instance
runs, which depends upon the quantity of resources assigned to it.
Parameters~$e_\text{p}(n)$ and~$e_\text{f}(l,n)$ express, respectively,
the power consumed at node~$n$ to provide one unit of CPU, and the power
consumed at that node to support the memory and storage requirements of
an instance of layer~$l$. Both quantities are parameters for our model
and can be set following the methodology introduced
in~\cite{mei2017energy}, i.e., analyzing the requirements of individual
layers and the resulting energy consumption. 
Furthermore, each instance~$(l,i)$ running at node~$\nu(l,i)\in\Nc$
implies additional energy consumption due to data transmissions over the
network:
\begin{equation}
\label{eq:enet}
E^{l,i}_\text{net}=e_\text{net}(\nu(l,i))\sum_{(m,j)\in\Ic}\chi(l,i,m,j).
\end{equation}
In \Eq{enet}, the energy spent for data transmissions for layer
instance~$(l,i)\in\Ic$ is given by the product of (i) a
factor~$e_\text{net}(\nu(l,i))$, expressing how much energy is required
by node~$n$ to transmit one unit of data, and (ii) the quantity of data
going from~$(l,i)$ to any other layer instance~$(m,j)$.

The energy consumed during one epoch can be found summing the
instance-specific energy consumption values \Eq{energy-v}: 
\begin{equation}
\label{eq:energy}
\Eb(x,y,z,\rho)=
\sum_{(i,l)\in\Ic}\left(E_\text{
comp}^{l,i}+E_\text{net}^{l,i}\right).
\end{equation} 
Finally, $\Eb(x,y,z,\rho)$ can be mapped into
{\em carbon} emissions following 
the methodology in~\cite{lacoste2019quantifying}, 
which accounts for the  quantity of~$CO_2$
emitted for each kilowatt-hour of consumed energy.

\subsection{Overall learning time and energy consumption}
\label{sec:accuracy}

In \Sec{timecost}, we have derived the time~$\Tb$ and energy~$\Eb$
required by each epoch as functions of our decision variables. 
In order to obtain the total time and energy required by the overall learning process, we need to multiply such values by the number $\Kb(y,\epsilon^{\max})$~of epochs needed to attain the target learning quality~$\epsilon^{\max}$. Deriving a closed-form expression for the number~$\Kb$, in a manner similar to~$\Tb$ and~$\Eb$,  
is currently possible only for few, simple
scenarios among those that we address. 
Indeed, as discussed in \Sec{relwork}, 
currently-available results only target scenarios where {\em all}
nodes share the same DNN, and {\em all} layers of said DNN are averaged,
i.e., {\em either} data {\em or} model parallelism are employed, but not
a combination of both. 
Thus, to show the ability of our approach to deal with arbitrary -- 
and potentially more efficient -- instance trees and assignment
decisions, we  use an auxiliary  ML model  to estimate~$\Kb$, owing to 
the ML ability to reveal and leverage the
structure underlying those phenomena that are too complex to describe 
through a traditional model.

A key observation we make is that, if we assume to always use all the
available data, i.e., to honor constraints \Eq{constr-flow} and
\Eq{constr-data} with the equality sign, then the number~$\Kb$ of epochs
to run only depends upon the instance tree we consider, i.e., the~$y(l,i,m,j)$
variables. 
Given such decisions, as well as the target learning quality~$\epsilon^{\max}$, 
our approach follows the steps set forth below:
\begin{enumerate}
    \item[{\em (i)}] we run a number~$M$ of experiments, each for different 
    values of the $y$~variables;
     \item[{\em (ii)}]  for each experiment, we determine the resulting value
of~$\Kb(y,\epsilon^{\max})$;
    \item[{\em (iii)}]  using the above information, we  train an {\em auxiliary} DNN
that can predict the value of $\Kb(y,\epsilon^{\max})$, given arbitrary values for the $y$ variables.
\end{enumerate}
A similar approach has been used, among others,
in~\cite{sim2018online} for mmWave-based vehicular networks, and
in~\cite{rottondi2018machine} for optical networks. 

The output of our experiments
is collected as a 5-dimensional tensor whose shape
is~$|\Lc|{\times}\alpha|\Dc|{\times}|\Lc|{\times}\alpha|\Dc|{\times}
M$, where $|\Lc|$~is the number of layers, $\alpha|\Dc|$~is the
maximum number of instances we can create for each layer, and~$M$
is the number of experiments we perform. For each
experiment~$\omega\in\{ 1\dots M\}$, the 
corresponding entry in the 5-dimensional tensor 
 contains the decisions~$y(l,i,m,j)$.
The auxiliary DNN gives as output 
the number~$\Kb(y,\epsilon^{\max})$ of epochs to run, in
order to achieve learning quality~$\epsilon^{\max}$.
 
To identify the best auxiliary DNN, 
we evaluate three  architectures, 
\begin{enumerate}
    \item the basic architecture (``MaxCNN'' in plots), with two
convolutional layers and two fully-connected ones, and a max-pooling
layer after each convolutional one;
    \item a variant thereof (``AvgCNN'' in plots), where average pooling
layers are used {\em in lieu} of max-pooling ones;
    \item a non-convolutional network (``FConly'' in plots), where both
convolutional layers are replaced by as many fully-connected ones, with
the same input and output sizes.
\end{enumerate}
\begin{figure}[b]
\centering
\includegraphics[height=3.25cm]{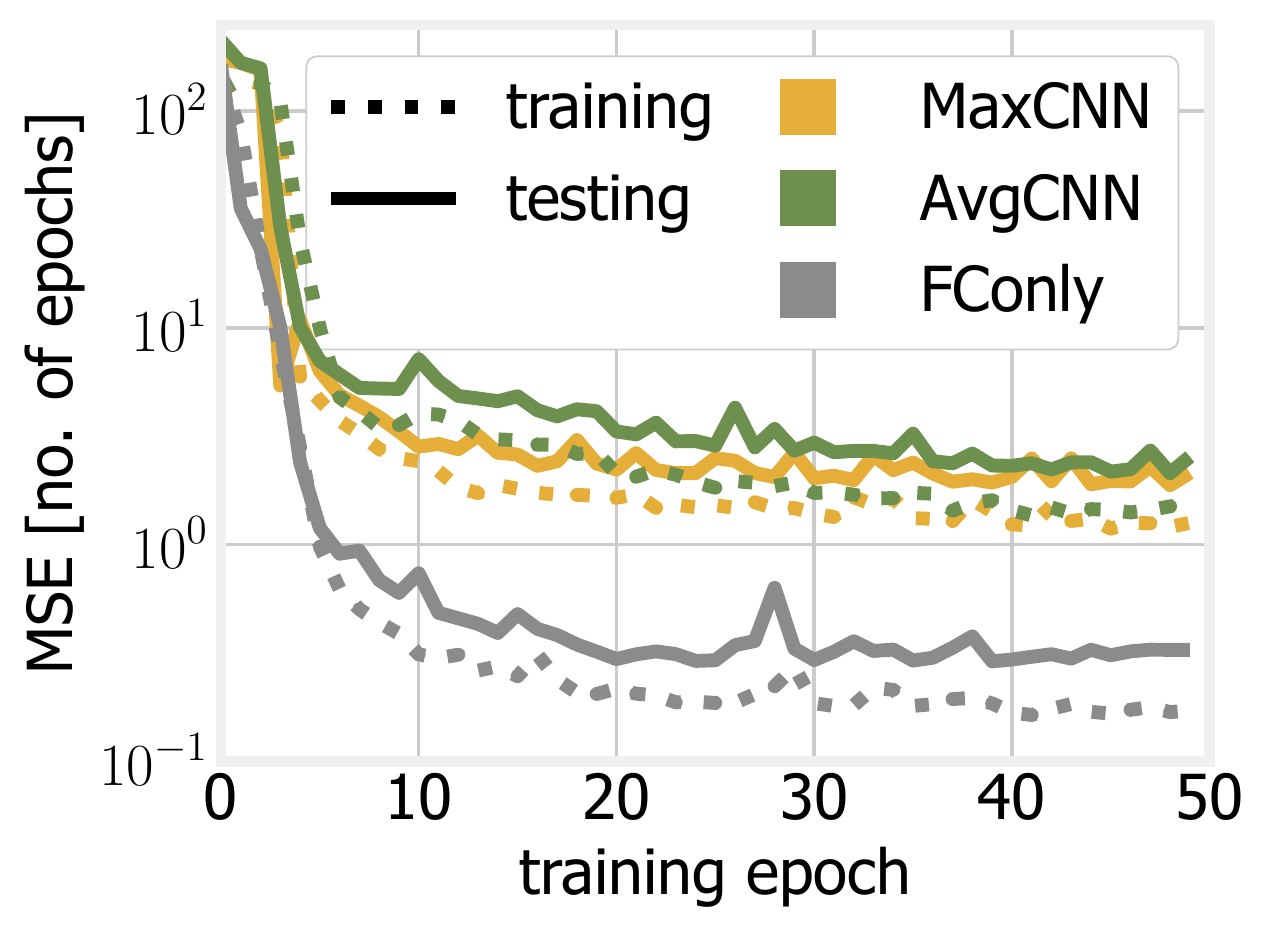}
\includegraphics[height=3.25cm]{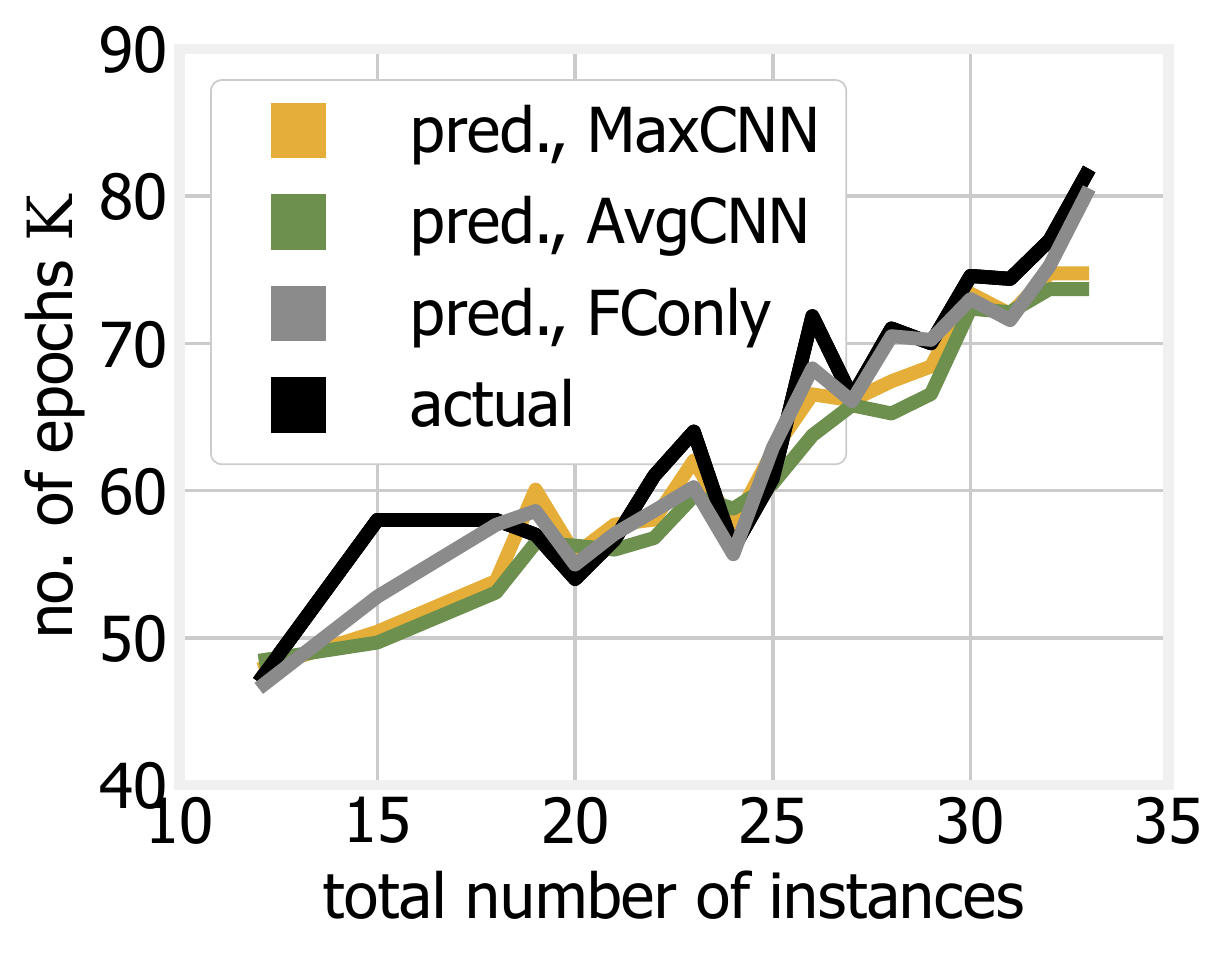}
\caption{
Auxiliary DNN for estimating the number of required epochs: 
resulting MSE (left), and real and predicted values of the number~$K$ of iterations (right).
    \label{fig:aux}
} 
\end{figure} 
In all cases, we are facing a regression problem, hence, we adopt the mean square error (MSE) as loss function.

We validate our approach using an image classification task leveraging the 
AlexNet DNN~\cite{krizhevsky2012imagenet} and the CIFAR-10~\cite{krizhevsky2009learning} dataset. 
The results are summarized in \Fig{aux}(left), depicting
the evolution of the MSE across training epochs.  
We can observe that,
for all the auxiliary DNN architectures, both training and testing MSE values rapidly
converge to very small values, below~$1$ for the FConly architecture.
This highlights how our approach is indeed very effective in predicting
the number~$K$ of iterations required for convergence. \Fig{aux}(right)
further shows that, consistently with~\cite{neglia}, $K$ tends to grow
as the number of layer instances increases, and the FConly architecture is
always associated with very close predictions.

Since the applicability of our methodology to real-world cases hinges on the
availability of sufficient training data,  
it is  worth highlighting that experiments like ours are routinely
performed upon evaluating and adopting a new DNN architecture, hence,
obtaining results similar to those in \Fig{aux}(right)
comes at a modest cost in terms of additional work. 
Moreover,   
transfer learning 
could be leveraged to further extend the applicability of the available experiments.

\section{The RightTrain Solution}
\label{sec:algo}

As proved in \Sec{analysis}, directly 
optimizing \Eq{obj} subject to constraints \Eq{st}, 
is a daunting task. We thus introduce a new, effective heuristic, 
called RightTrain, which {\em decouples} the decisions of 
(i) choosing the instance tree, (ii) performing instance-to-node mapping, and (iii) 
assigning the necessary resources.
At every step, {\em efficiency} is the main criterion driving RightTrain decisions.

\begin{figure}
\centering
\includegraphics[width=1\columnwidth]{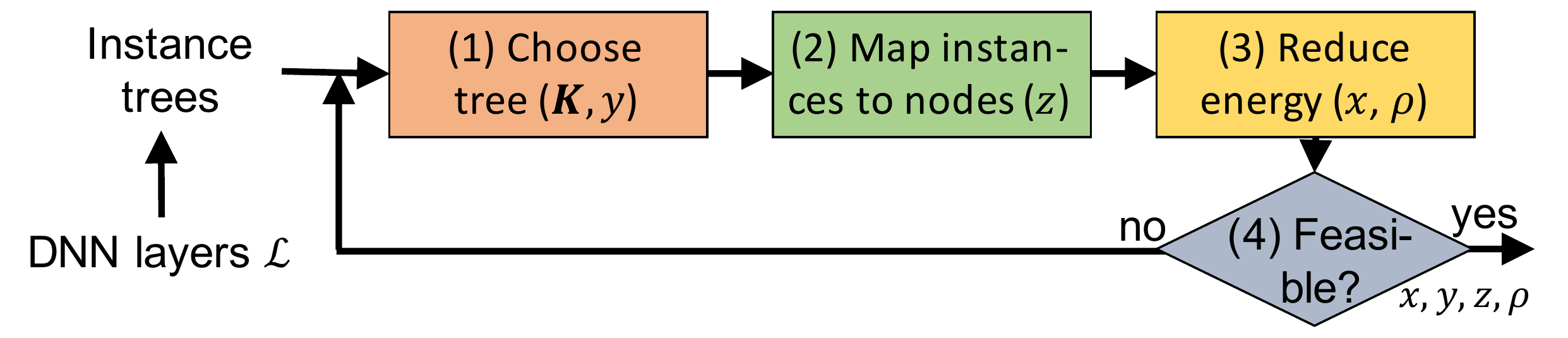}
\caption{
Main steps of the RightTrain solution concept: given the set of layer instance trees to consider, 
RightTrain selects, at each iteration, the hitherto-untested tree associated with the lowest 
energy consumption (Step~1, \Sec{sub-trees}). For such a tree, 
it makes the near-optimal layer instance-to-node mapping decisions (Step~2, \Sec{sub-mapping}), 
and further improves efficiency by tweaking data and resource utilization (Step~3, \Sec{sub-refine}).
    \label{fig:flowchart}
}
\end{figure}

As summarized in \Fig{flowchart}, RightTrain takes as an input the set of instance trees 
 to consider (like the one in \Fig{fresco2}); such a set can be efficiently computed offline, 
 in a scenario- and application-dependent manner. 
 RightTrain then iterates over the set of instance trees, selecting at each step the one 
 requiring the least amount of {\em total} processing (Step~1 in \Fig{flowchart}, 
 detailed in \Sec{sub-trees}). For each tree, the $y$-variables are fixed, hence, 
 in Step~2 (\Sec{sub-mapping}) we make the mapping decisions~$z$, under the (temporary) assumption 
 that (i) all the data of the selected sources is used, and (ii) all the processing 
 capabilities at each node are exploited. Both assumptions are dropped 
 in Step~3 (\Sec{sub-refine}), which seeks to refine the solution obtained in Step~2 
 by using less data and/or less computing power, thereby reducing the energy consumption 
 without jeopardizing the learning performance. 
If a feasible solution is obtained, then the algorithm terminates (Step~4); otherwise, 
it goes back to Step~1 and moves to the next instance tree.

\subsection{Layer instance tree ordering}
\label{sec:sub-trees}

Step~1 of the RightTrain solution requires choosing, from the set of layer instance trees 
to consider, the next one to try. Ideally, we would like to select a tree minimizing the 
energy consumption \Eq{obj}, however, this is not possible as instance-to-node mapping and 
resource assignment decisions (respectively, Steps~2 and~3) have yet to be made. 
Nonetheless, the instance tree -- along with information on layer and data source characteristics -- 
allows for estimating the total quantity of {\em processing} entailed by the tree itself. 
Specifically, recalling that~$\Ic$ is the set of layer instances created for a given tree 
(i.e., DNN structure) and that $y$-decisions represent the tree topology, 
we can express the amount of required processing as: 
\begin{equation}
\label{eq:processing}
\Kb(y,\epsilon^{\max})\sum_{l\colon (l,i)\in\Ic} 
r(l)\hspace{-3mm}\sum_{\substack{d\in\Dc\colon (l,i)\text{ is}\\\text{an ancestor of }d}}\hspace{-3mm}\Delta(d)\hspace{-3mm}\prod_{\substack{m\in\Lc\text{ in path}\\ \text{from }d\text{ to }l}}\hspace{-3mm}q(m).
\end{equation} 
Looking at\,\Eq{processing} from right to left, the processing required by a given layer of a DNN 
for each epoch depends upon~\cite{green-ai}: 
(i) the quantity of data it processes
(which in turn depends upon the $q$-coefficients of the layers traversed before~$l$) 
and (ii) the layer complexity. 
Such a quantity is then summed across all layer instances, and multiplied by the 
number~$\Kb(y,\epsilon^{\max})$ of epochs to run before convergence.

In addition to being a sound criterion to follow in all cases, 
selecting layer instance trees associated with  low  processing load \Eq{processing} often results, 
as proved in \Sec{analysis}, in selecting trees yielding a low value of the  objective \Eq{obj}. 
Indeed, energy consumption \Eq{energy} is often dominated by the processing energy \Eq{energy-v}, which 
in turn depends upon three quantities also accounted for in \Eq{processing}, 
namely, the number~$\Kb$ of iterations, the layer complexity~$r(l)$, and 
the quantity of available data~$\Delta$. 

\subsection{Layer instance-to-node mapping}
\label{sec:sub-mapping}

\begin{figure}
\centering
\includegraphics[width=1\columnwidth]{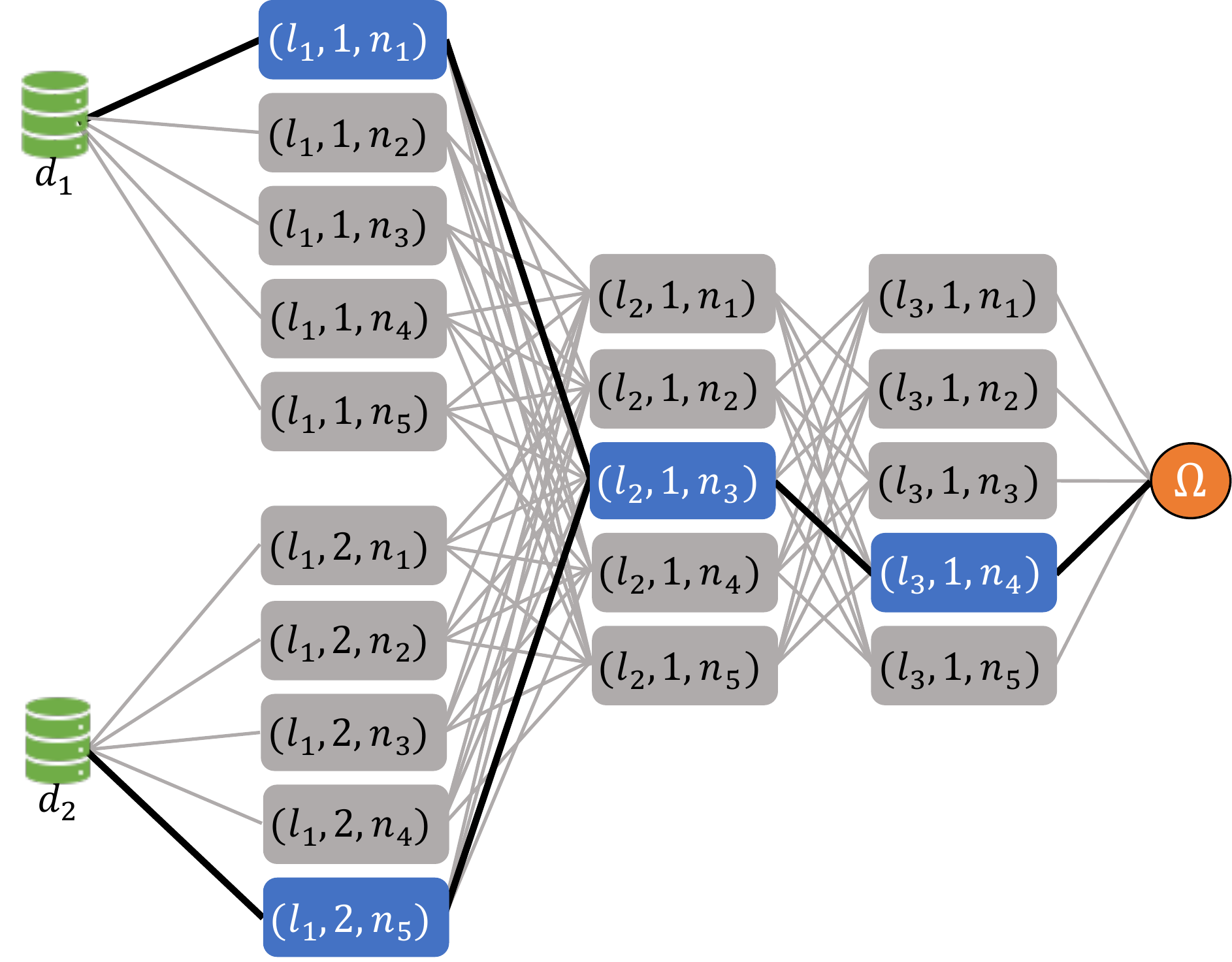}
\caption{
    The expanded graph representing the possible decisions in \Fig{fresco}, 
    with colored nodes and bold edges highlighting the DA-ST corresponding to mapping decisions in \Fig{fresco3}.
    \label{fig:bigtree}
} 
\end{figure}

Step~3 of RightTrain  {\em maps} layer instances in~$\Ic$ to nodes in~$\Nc$, i.e., it decides 
at which node each layer instance should run. As mentioned above, initially such decisions are made 
under the assumption that all available data and computational capabilities are used. 
This problem is combinatorial and, in general, hard to approach. On the positive side, 
however, we can leverage the {\em tree} structure connecting layer instances, 
as in \Fig{fresco2}. Specifically, our approach is to (i) build the {\em expanded graph} 
shown in \Fig{bigtree}, summarizing all possible mapping decisions, and (ii) 
build a delay-aware Steiner tree (DA-ST) upon such a graph.  
The DA-ST is the minimum-weight tree spanning a given subset of the vertices of an undirected graph, 
called {\em terminal vertices}, with the additional constraint that the maximum learning 
time~$T^{\max}$ must be honored, as per 
the $T^{\max}$-clause in \Eq{st}.

The expanded graph is built as follows:
\begin{enumerate}
    \item it contains a vertex $(l,i,n)$ for each possible deployment
decision, i.e., for each layer instance-node pair such that~$\mu(l,i,n)=1$;
    \item an edge is drawn between vertices~$(l,i,n)$ and~$(m,j,n')$ if 
    the layer instances are connected in the instance graph and 
     the nodes~$n$ and~$n'$ can communicate, i.e., if~$y(l,i,m,j)=1
~\wedge ~S(n,n')>0$;
    \item we also create an additional vertex for each data source
in~$d\in\Dc$, and connect such a vertex to all vertices 
    in the expanded 
    graph representing layer instances to which $d$~is connected.
    \item we add a further vertex~$\Omega$ connected to all
vertices~$(l_L,i,n)$ corresponding to the last layer~$l_L$;

    \item the weight of each edge~$((l,i,n),(m,j,n'))$ corresponds to
the energy consumption, as per \Eq{energy-v}, due to 
    the deployment decision represented by vertex~$(m,j,n')$, i.e.,
running instance~$(m,j)$ at node~$n'$.
\end{enumerate}

\begin{algorithm}[t]
\caption{Layer instance-to-node mapping 
    \label{alg:mapping}}
\begin{algorithmic}[1]
\Require{Expanded graph~$\{d\}\cup\{(l,i,n)\}\cup\{\Omega\}$} \label{line:given}
\State{$\Tc\gets\{\Omega\}$} \label{line:init-t}
\While{$\Dc\setminus\Tc\not\equiv\emptyset$} \label{line:while}
 \State{$w^\star,\pi^\star\gets\infty,\emptyset$} \label{line:init-vars}
 \ForAll{$d\in\Dc\setminus\Tc,v\in\Tc$}
  \State{$w,\pi\gets\mathsf{RestrictedMinWeightPath}(d,v)$} \label{line:get-path}
  \If{$w<w^\star$}
   \State{$w^\star,\pi^\star\gets w,\pi$} \label{line:best-path}
  \EndIf
 \EndFor
 \State{$\Tc\gets\Tc\cup \pi^\star$} \label{line:add-best}
\EndWhile
\ForAll{$v=(l,i,n)\in\Tc$}
 \State{$z(l,i,n)\gets 1$} \label{line:setz}
\EndFor
\State\Return{$\{z\}$} \label{line:retz}
\end{algorithmic}
\end{algorithm}

As exemplified in \Fig{bigtree}, 
the DA-ST connects the vertices representing the data sources with~$\Omega$, 
and covers the same topology as the layer instance tree. Since the DA-ST is the {\em minimum-weight} 
among the trees with these features, its vertices 
represent the layer instance-to-node mapping that minimizes energy consumption (see \Eq{obj}),
under the aforementioned conditions, i.e., that all data and computation resources are used. 
The mapping decisions are made as summarized in \Alg{mapping}. Given the expanded graph 
whose vertices are (i) data sources in~$\Dc$, (ii) vertices of type~$(l,i,n)$ representing 
possible mappings, and (iii) special vertex~$\Omega$ (\Line{given}), we  build 
the DA-ST~$\Tc$.
In \Line{init-t}, we initialize the tree~$\Tc$, so as to include vertex~$\Omega$. 
Then, so long as there are data sources not yet included in the tree (\Line{while}), 
we look for the minimum-weight path such that (i) it connects a data source~$d$ not yet in the 
tree with a vertex~$v$ of the DA-ST itself, and (ii) does not break constraints \Eq{st}. 
To this end, in \Line{init-vars} the minimum weight~$w^\star$ and the minimum-weight 
path~$\pi^\star$ are initialized to~$\infty$ and~$\emptyset$, respectively. 
Then function~$\mathsf{RestrictedMinWeightPath}(d,v)$ (\Line{get-path}) provides 
the path~$\pi$ connecting each data source~$d$ not yet reached by the tree with each vertex~$v$ 
already in the tree. 
The minimum-weight path is then identified (\Line{best-path}) and added to 
the DA-ST~$\Tc$ in \Line{add-best}. 
Once all data sources have been included, the DA-ST is completed. 
The algorithm therefore sets to~$1$ the $z$-variable corresponding to the selected DA-ST vertices 
(\Line{setz}), and returns them.

The procedure~$\mathsf{RestrictedMinWeightPath}$ uses the algorithm proposed 
in~\cite{lorenz2001simple} to find the minimum-weight path connecting~$v$ and~$d$ 
while honoring delay requirements. Finding such a path requires 
solving an instance of the {\em constrained shortest path} problem. 
The problem itself is NP-hard, however, the heuristic~\cite{lorenz2001simple} can solve it 
within $\epsilon$ ($\epsilon\geq 0$) from the optimum in polynomial time.

As proven in \Sec{analysis}, \Alg{mapping} as a whole has polynomial 
(namely, $O(\left|\Ic|^3|\Nc|^3\frac{1}{\epsilon}\right)$) time
complexity has a constant competitive ratio, namely, it is
within~$\left(2-\frac{2}{W}\right)(1+\epsilon)$ from the optimum,
where~$W$ is the number of vertices of the optimal DA-ST.

\subsection{Decisions refinement}
\label{sec:sub-refine}

In Step 2, we have decided
which layer instances to create and how to connect them, i.e., the layer instances in~$\Ic$ and the edges in the layer instance 
tree connecting them, as expressed through the $y$-variables, 
and how to map layer instances to physical nodes, i.e., the $z$-variables. The values of both  
variable sets have been obtained under the assumption that all data from the selected data sources and 
all the capabilities of physical nodes are used. 
In the spirit of recent efficiency-focused research~\cite{green-ai}, Step~3 
seeks to establish whether {\em all}  
that data and that computational power 
is really needed. 
Our goal is thus to obtain a solution that meets
all constraints in \Eq{st}, including the minimum learning quality and maximum time, 
while further improving the energy objective \Eq{obj}.

Given the~$y$- and $z$-variables, the problem of optimizing \Eq{obj} subject to constraints 
\Eq{st} only has continuous variables, 
namely,~$x$ and~$\rho$. It follows that such a problem can be efficiently solved:
(i) by off-the-shelf solvers, e.g., CPLEX or Gurobi, if a closed-form expression 
    is available for $\Kb$, $\Eb$ and~$\Tb$, or (ii) 
through iterative, gradient-based methods like BFGS~\cite{fletcher2013practical},
    if such closed-form expressions are not available. 
Even more importantly, as formally proven in \Sec{analysis}, such a continuous problem 
is {\em convex} in many practical cases. It follows that numerical approaches 
(either off-the-shelf solvers or gradient-based methods) are {\em guaranteed} 
to find the optimal values of~$x$ and~$\rho$ with very good efficiency -- 
polynomial worst-case complexity~\cite{boyd2004convex}, and much faster than that in most cases.

\section{Problem and Algorithm Analysis}
\label{sec:analysis}

In this section, we prove several important properties of the problem we solve and the RightTrain approach.
We starting from showing that the problem is  NP-hard.
\begin{property}[Problem hardness]
\label{prope:hard}
The problem of optimizing \Eq{obj} subject to constraints 
\Eq{st} is NP-hard.
\end{property}
\begin{IEEEproof}
We prove the thesis through a reduction from the Generalized Assignment 
Problem~\cite{cattrysse1992survey} (GAP), requiring to assign a set of {\em tasks} to a set of {\em agents}; 
each task-to-agent 
assignment incurs a given {\em cost}, and the goal is to minimize the total cost. 
More specifically, we reduce instances of the GAP to {\em simpler} instances of our own problem, where:
\begin{itemize}
    \item there is only one data source~$\Dc=\{d\}$;
    \item the layer instance graph is a chain, the redundancy factor is~$\alpha=1$ and $q(l)=1$ 
    for all layers, i.e., there is only one instance per layer and the same traffic traverses them all;
    \item the number of iterations goes to~$\infty$ if any $x$-value is lower than~$\Delta(d)$, 
    i.e., we must use all data;
    \item the number of iterations goes to~$\infty$ if any layer instance is assigned less 
    than a quantity $\rho_0^l$ of computational capabilities, and 
the timeout~$T^{\max}$ is set to~$\infty$ -- hence, it is optimal to always set $\rho(l,i)=\rho_0^l$;
    \item communication links have infinite capacity and zero delay.
\end{itemize}
Due to the conditions listed above, the values of $y$, $x$~and $\rho$~variables can be 
trivially set, and our problem reduces to instance-to-node mapping, i.e., setting the $z$-variables.

Therefore, we can reduce an instance of the GAP problem to an instance of the simplified problem 
we stated above by:
\begin{enumerate}
    \item creating one layer (hence, one layer instance) per task;
    \item creating one node per agent;
    \item setting the fixed energy consumption~$e_\text{f}(l,n)$ equal to the cost of assigning 
    task~$l$ to agent~$n$.
\end{enumerate}
This implies that an instance of a known NP-hard problem, namely, GAP, can be reduced 
to an instance of ours in polynomial (indeed, linear) time, which 
proves the thesis.
\end{IEEEproof}
We remark that the proof of \Prope{hard} reduces GAP instances to greatly 
simplified instances of our own problem; this allows us to conjecture that our problem, 
besides being NP-hard, is also significantly more complex than an NP-hard problem like GAP.

Let us now move to the RightTrain heuristic, and focus on Step~1 therein, i.e., the choice of the 
next layer instance graph to consider. In \Sec{sub-trees}, we formulated a selection 
criterion based on \Eq{processing}, expressing the quantity of processing associated with 
a given tree.
Next, we prove that such criterion often results in selecting the instance tree with the lowest energy consumption. 
\begin{property}[Processing and energy]
\label{prope:powerenergy}
If proportional energy factors are the same for all usable nodes and dominate the global energy consumption, 
then a layer instance tree minimizing the quantity of processing \Eq{processing} also minimizes the objective \Eq{obj}.
\end{property}
\begin{IEEEproof}
The objective \Eq{obj} requires minimizing energy, which is given in
\Eq{energy}. 
Neglecting $e_\text{net}(n,n')$ and~$e_\text{f}(n,l)$, 
then \Eq{energy} reduces
to~$\sum_{(l,i)\in\Ic}t_\text{comp}^{l,i}e_\text{p}(\nu(l,i))\rho(l,i)$
which, 
recalling \Eq{time-one-comp} and considering the conditions stated in
\Sec{sub-trees} and that we are not making any mapping decision, 
the expression becomes~$\sum_{(l,i)\in\Ic}
e_\text{p}(\nu(l,i))r(l)\sum_{\substack{d\in\Dc\colon (l,i)\text{
is}\\\text{an ancestor of }d}}\Delta(d)$. 
Considering~$e_\text{p}(n)=e_\text{p}\, \forall n$ , 
we can re-write \Eq{obj} as: 
\begin{equation}
\nonumber
e_\text{p}
\sum_{(l,i)\in\Ic} r(l)\sum_{\substack{d\in\Dc\colon (l,i)\text{
is}\\\text{an ancestor of }d}}\Delta(d)\prod_{\substack{m\in\Lc\text{ in path}\\ \text{from }d\text{ to }l}}q(m),
\end{equation} 
which is exactly $\frac{e_\text{p}}{\Kb(y,\epsilon^{\max})}$~times the
quantity in \Eq{processing}.
\end{IEEEproof}
We remark that \Prope{powerenergy} describes very well scenarios where
layer instances are implemented within, e.g., containers, 
hence, with very small overhead.

Moving to  Step 2 of RightTrain, we show that
\Alg{mapping} has 
polynomial complexity, and a very good competitive ratio. 
\begin{property}[Complexity of \Alg{mapping}]
\label{prope:complexity}
\Alg{mapping} has a worst-case time complexity
of~$O\left(|\Dc||\Ic|^3|\Nc|^3\frac{1}{\epsilon}\right)$.
\end{property}
\begin{IEEEproof}
As shown in~\cite{lorenz2001simple}, the algorithm implementing 
the $\mathsf{RestrictedMinWeightPath}$~procedure 
has~$O\left(mn(\log\log n\left(1+\frac{1}{\epsilon}\right)\right)$ time
complexity, 
where $m$ and~$n$ are, respectively, the number of vertices and edges of
the input graph. 
In our case, the number of vertices of the expanded graph is
$O(|\Ic||\Nc|)$, and 
the number of its edges is $O(|\Ic|^2|\Nc|^2)$. The whole procedure is
repeated at 
most~$|\Dc|$ times, hence, the thesis follows.
\end{IEEEproof}
\begin{property}[Competitive ratio of \Alg{mapping}]
\Alg{mapping} has a competitive ratio of~$2(1+\epsilon)$.
\label{prope:compratio}
\end{property}
\begin{IEEEproof}
The structure of \Alg{mapping} mimics that of the algorithm 
in \cite{takahashi1980approximate} to solve the Steiner tree problem, which has a competitive 
ratio of~$2-\frac{2}{W}\leq 2$, where~$W$ is the number of vertices in the optimal Steiner tree. 
However, \Alg{mapping} contains a further source of suboptimality, namely, the 
$\mathsf{RestrictedMinWeightPath}$~procedure; as per \cite{lorenz2001simple}, 
its result is guaranteed to be within~$(1+\epsilon)$ from the optimum. 
Combining the two competitive ratios, the thesis follows.
\end{IEEEproof}

As for Step~3 of the RightTrain approach, it requires setting the~$x$ and~$\rho$ variables 
so as to further improve the energy objective \Eq{obj}, without jeopardizing the constraints in\,\Eq{st}. 
By leveraging theoretical arguments and experimental observations, 
we can state the following result.
\begin{proposition}[Convexity of the problem in Step~3]
\label{propo:convex}
The problem of optimizing \Eq{obj} subject to constraints
in \Eq{st}, with the~$y$- and $z$-values fixed, is convex.
\end{proposition}
The arguments supporting \Propo{convex} can be summarized as follows. 
Constraints \Eq{constr-oneinstance}--\Eq{constr-data} are clearly linear
in the variables~$x$ and~$\rho$, 
once~$y$ and~$z$ are given. Similarly, constraints
\Eq{time-one-net}--\Eq{energy} 
 reduce to linear expressions in variables~$x$ and~$\rho$. Also,
\Eq{time-one-comp}  
is convex in~$\rho(l,i)$, as it is easy to verify that its second
derivative is always positive.
As for the number of iterations needed for convergence, although no
closed-form expression for~$\Kb$ is available, 
theoretical and experimental
works~\cite{alaa,perlich2003tree,distributedQlearning} all concur that
the relationship 
between the quantity of used data and the resulting learning
quality (e.g., accuracy) 
is best captured by logarithmic functions, which are convex.

\section{Performance Evaluation}
\label{sec:peva}

After introducing our reference scenarios, in this section we evaluate the performance of  
RightTrain against split learning (SL) and the optimum (when the scenario size allows it).

\subsection{Reference scenarios}

\begin{figure}
\centering
\includegraphics[width=.9\columnwidth]{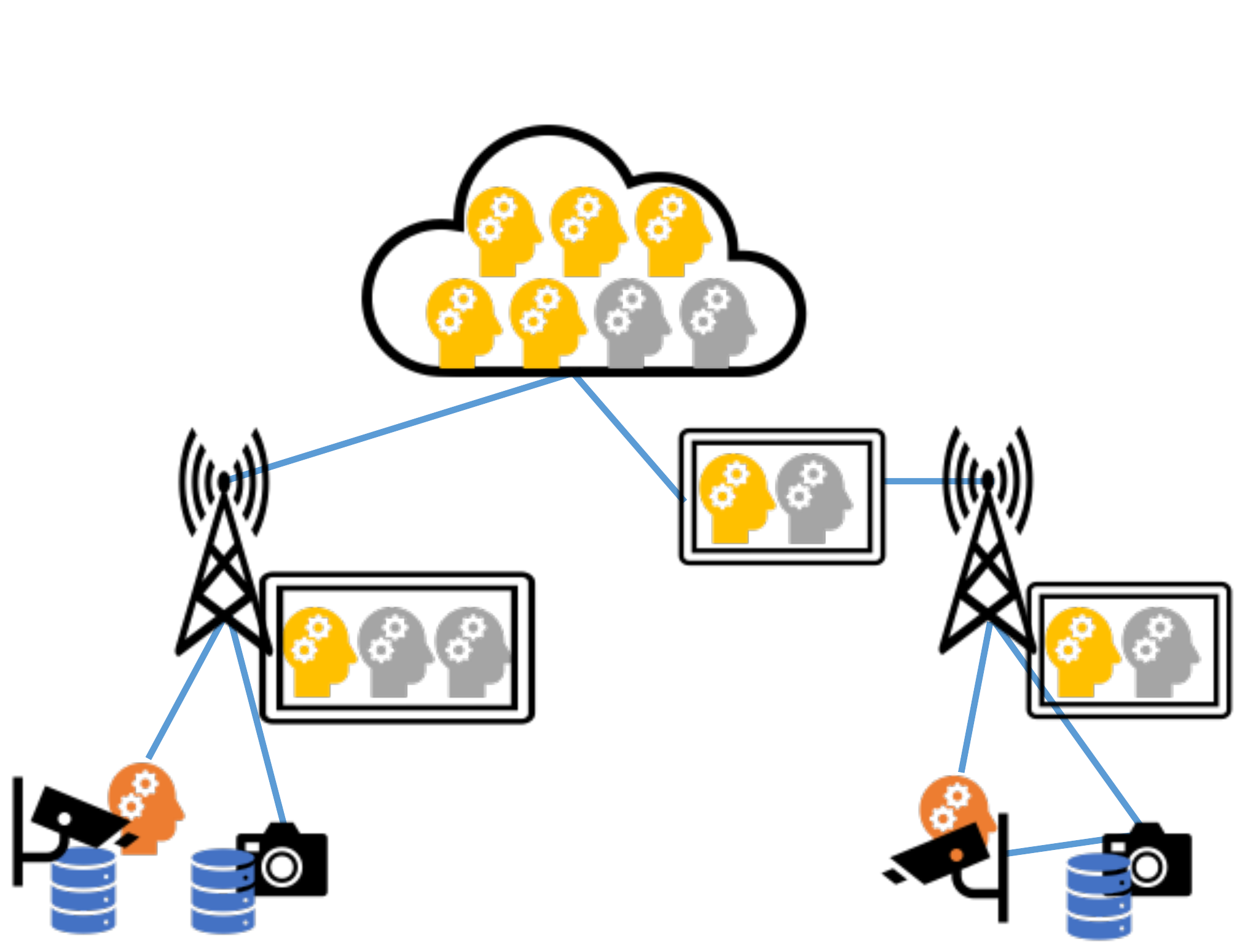}
\caption{
An example three-layered scenario, including mobile, edge, and cloud nodes. 
Brains denote different devices with computational capability, with the color of the 
brain corresponding to the category of the device itself (gold, silver, or bronze). 
Dark-blue cylinders denote data sources; light-blue edges connect pairs of devices that can communicate.
\label{fig:brains}
} 
\end{figure}

{\bf Learning task and DNN.} 
We consider an image classification task over the CIFAR
dataset, using a version of the AlexNet
DNN~\cite{krizhevsky2012imagenet}, including five convolutional layers
and three fully-connected ones. \Tab{layers} summarizes the layers
composing the AlexNet DNN, along with their complexity (per sample),
expressed in millions of operations (MOPs) per sample. Both CIFAR and
AlexNet are well-known, widely available and well-studied; this makes
our results more significant and easier to reproduce and generalize.

\begin{figure*}
\centering
\includegraphics[width=.32\textwidth]{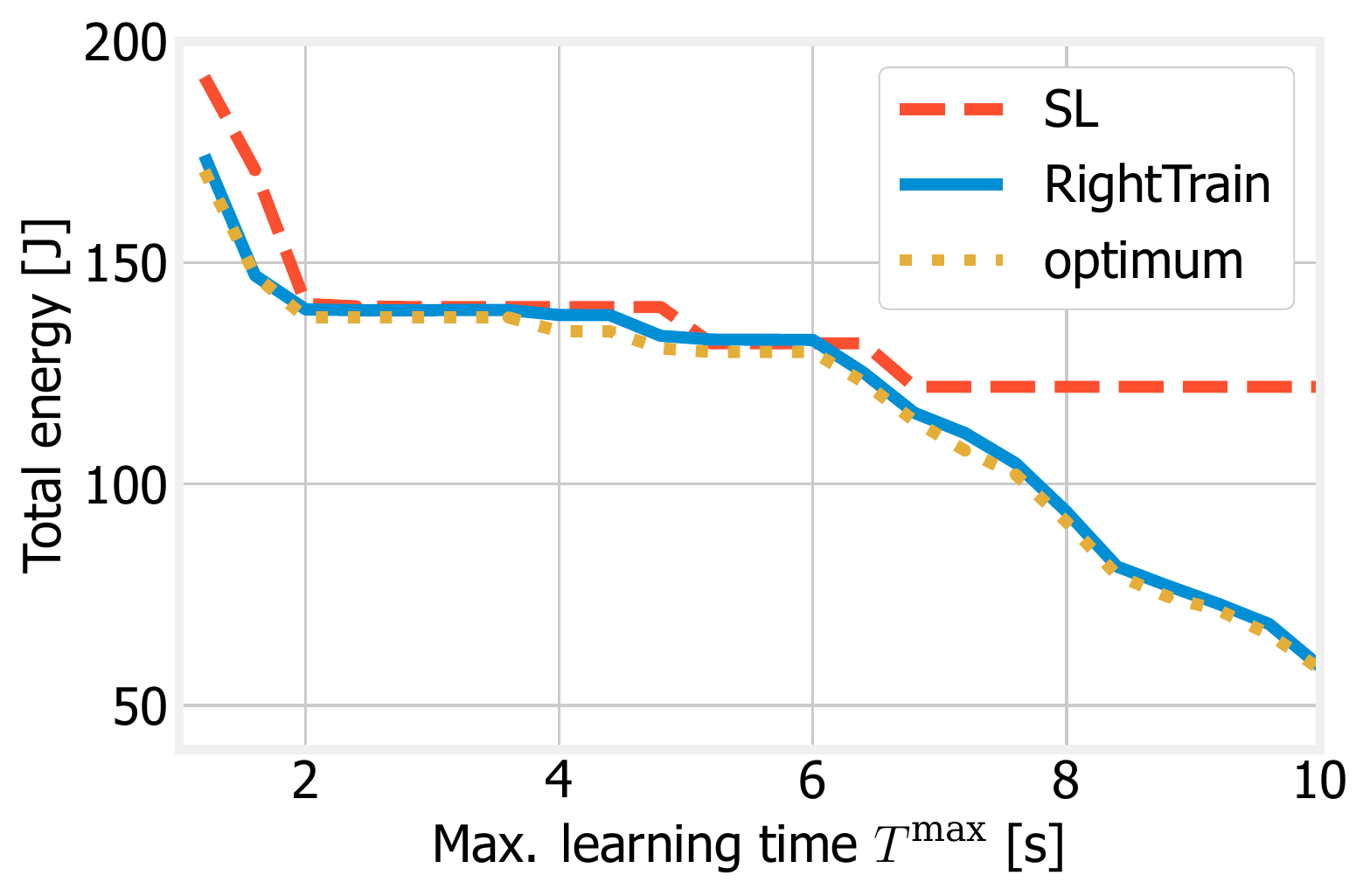}
\includegraphics[width=.32\textwidth]{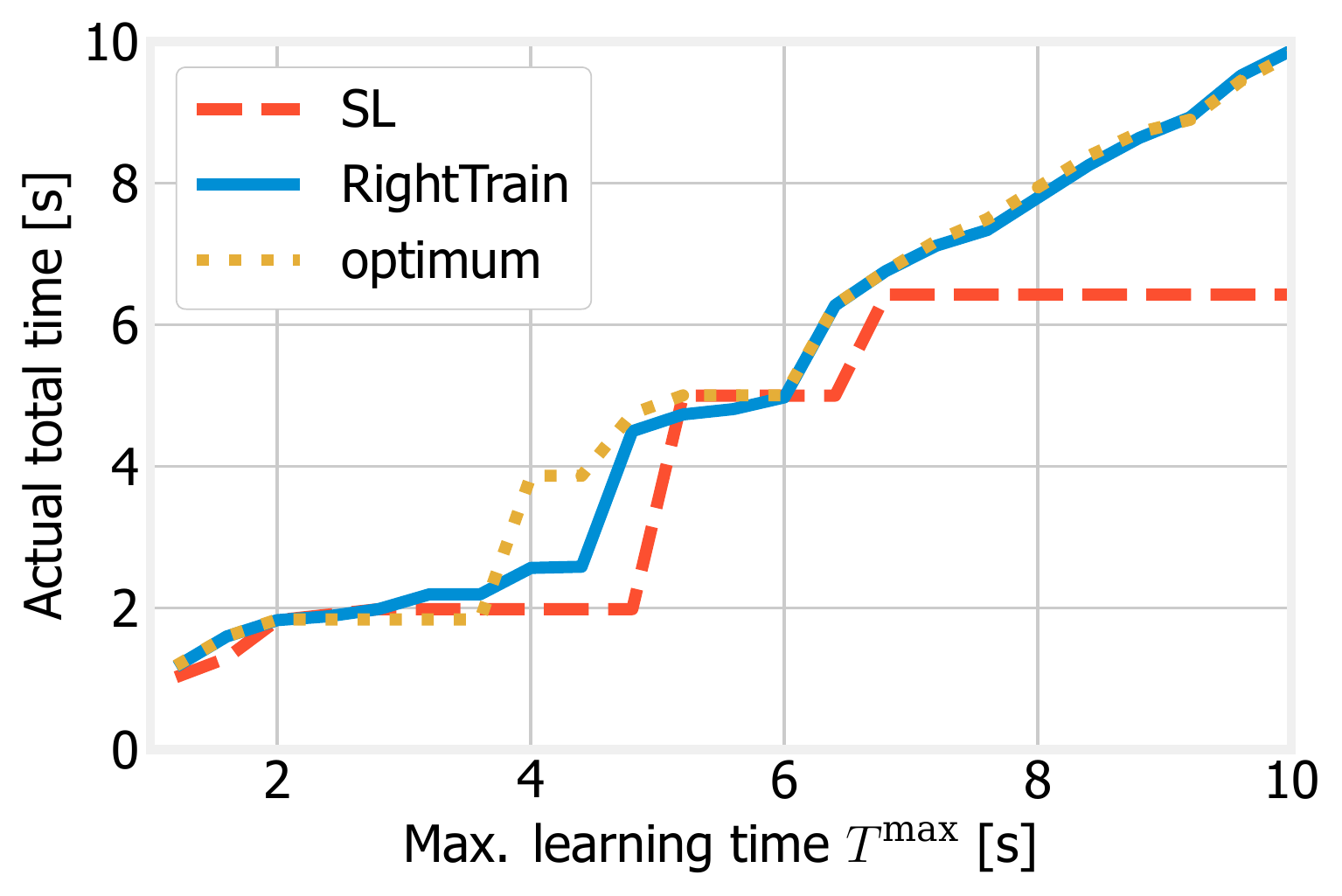}
\includegraphics[width=.32\textwidth]{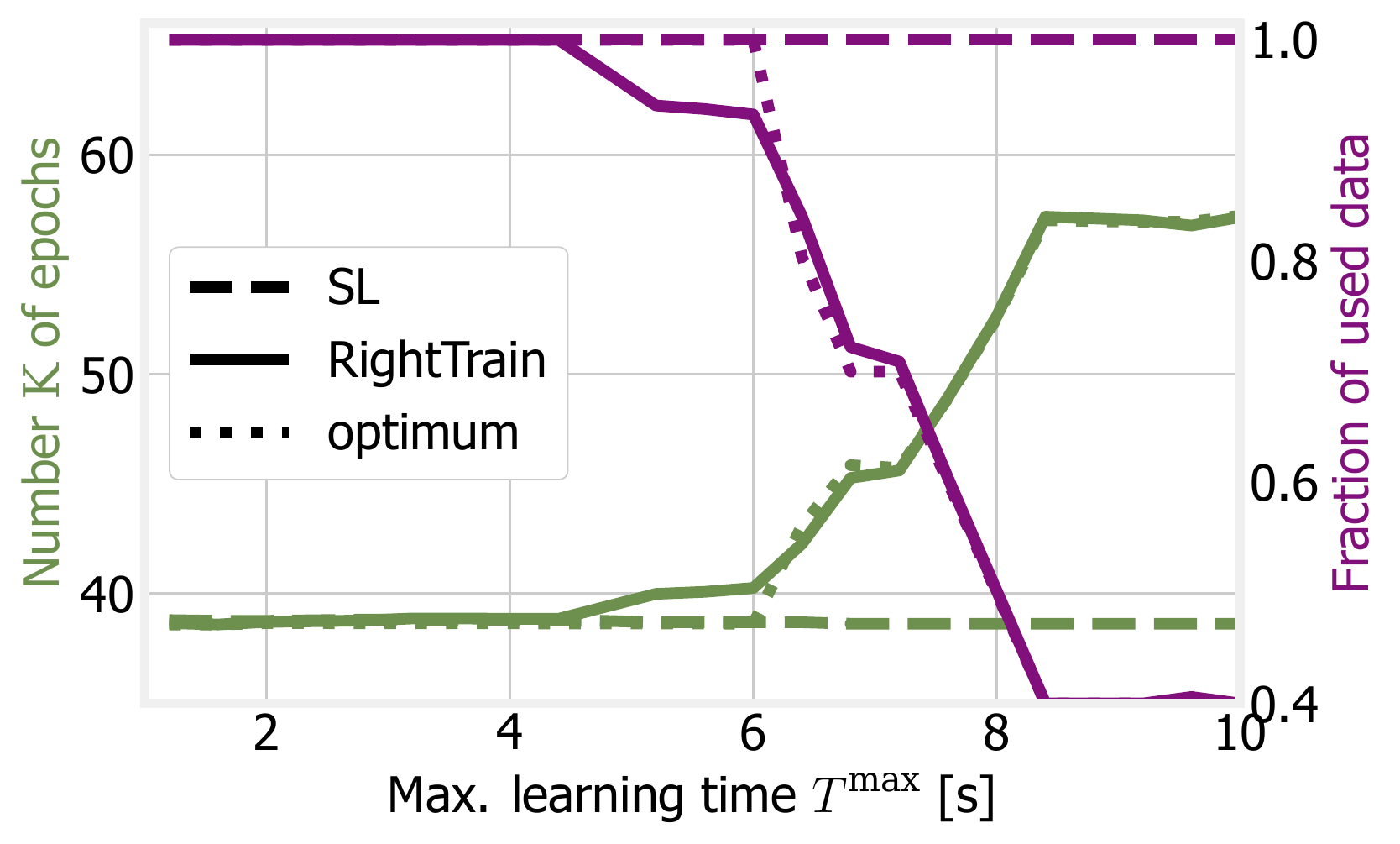}
\caption{
    Small-scale scenario: energy consumed as a function of the maximum learning time~$T^{\max}$ (left), actual and maximum learning time (center), number of iterations and fraction of used data (right).
\label{fig:small-general}
} 
\includegraphics[width=.32\textwidth]{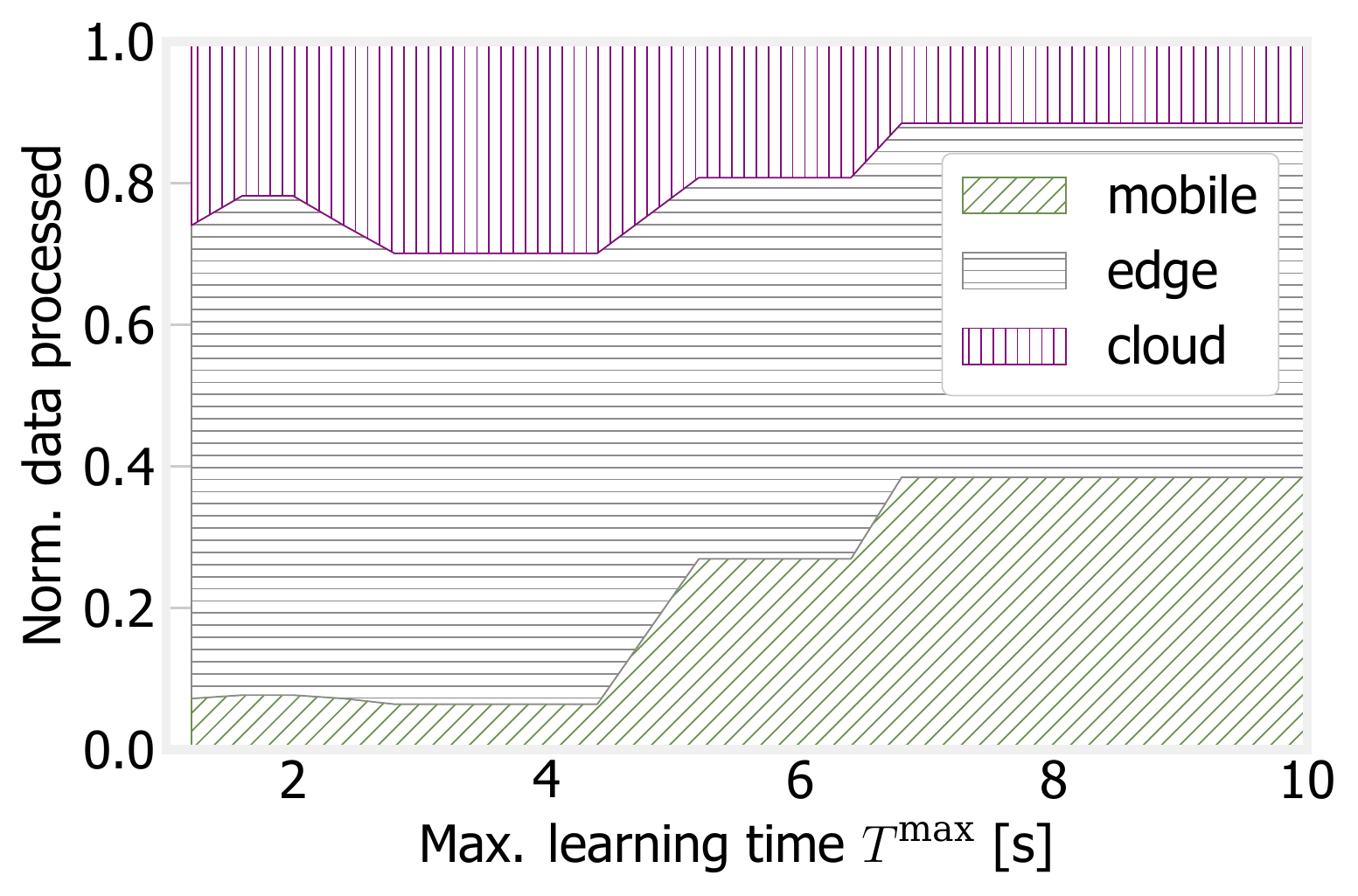}
\includegraphics[width=.32\textwidth]{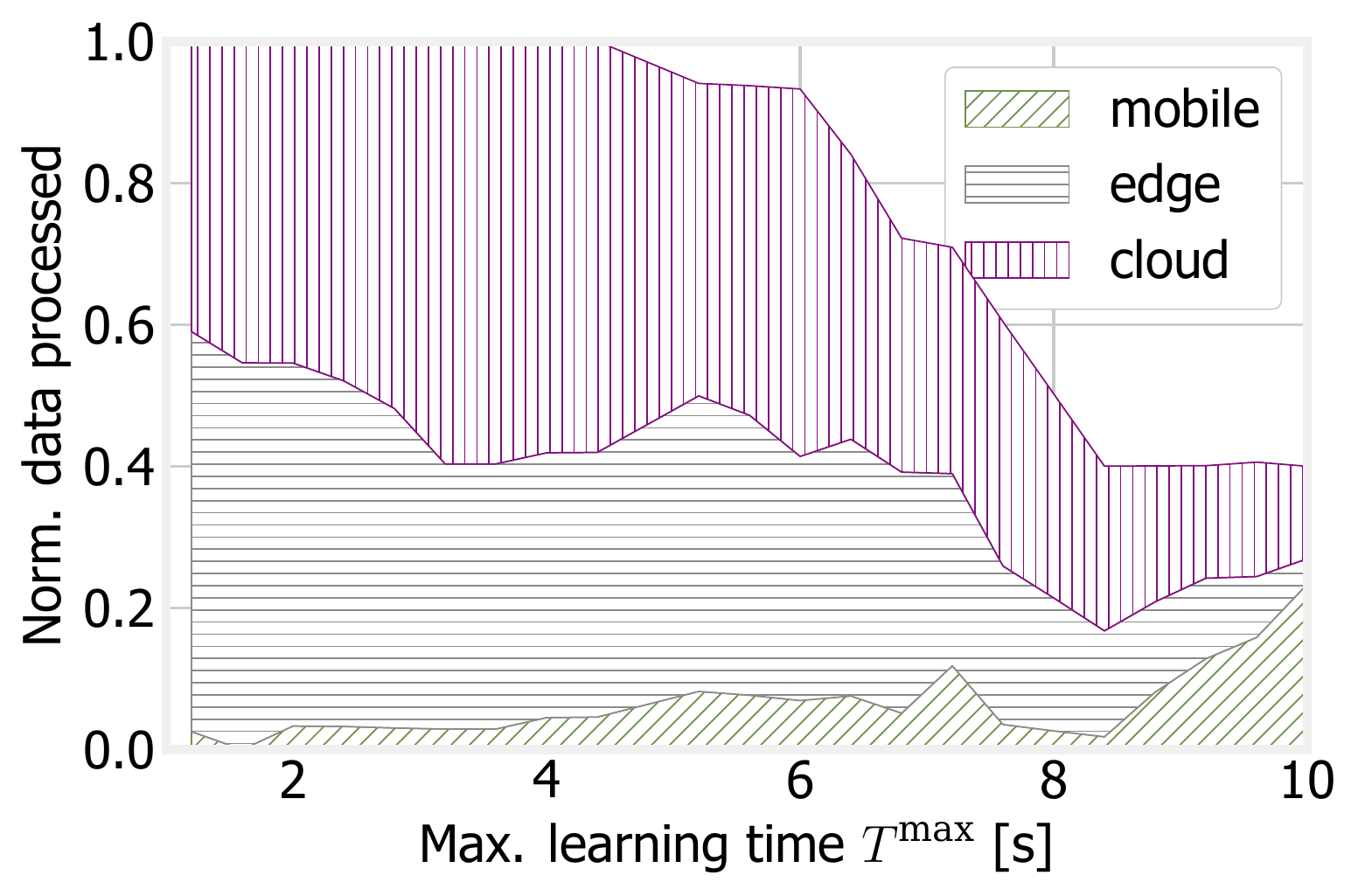}
\includegraphics[width=.32\textwidth]{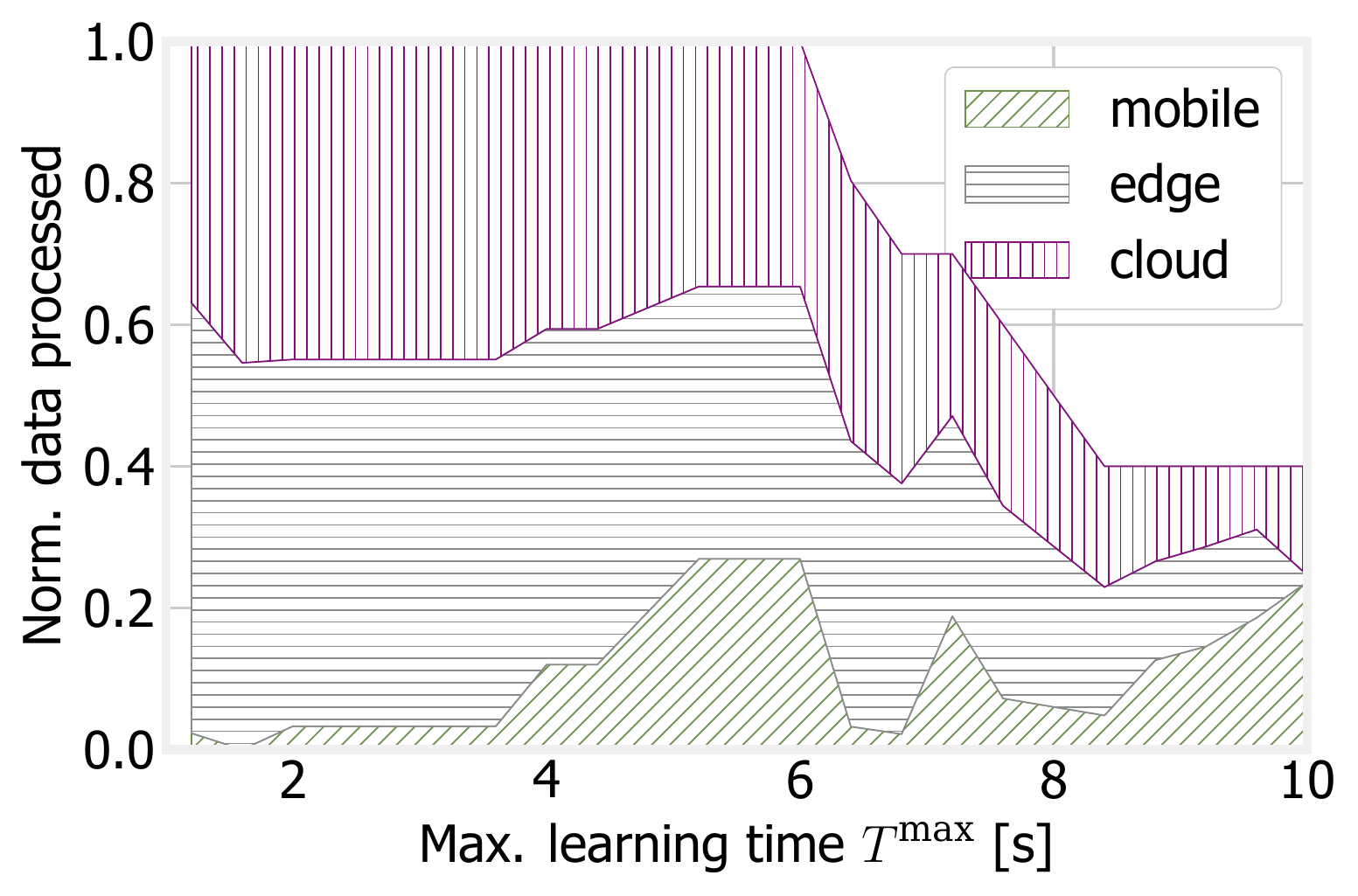}
\caption{
    Small-scale scenario: quantity of data processed at different parts of the network topology as the maximum learning time~$T^{\max}$ varies, under the SL (left), RightTrain (center), and optimal (right) strategies.
\label{fig:small-whereprocess}
} 
\end{figure*}

\begin{figure}[b!]
\centering
\includegraphics[width=.32\textwidth]{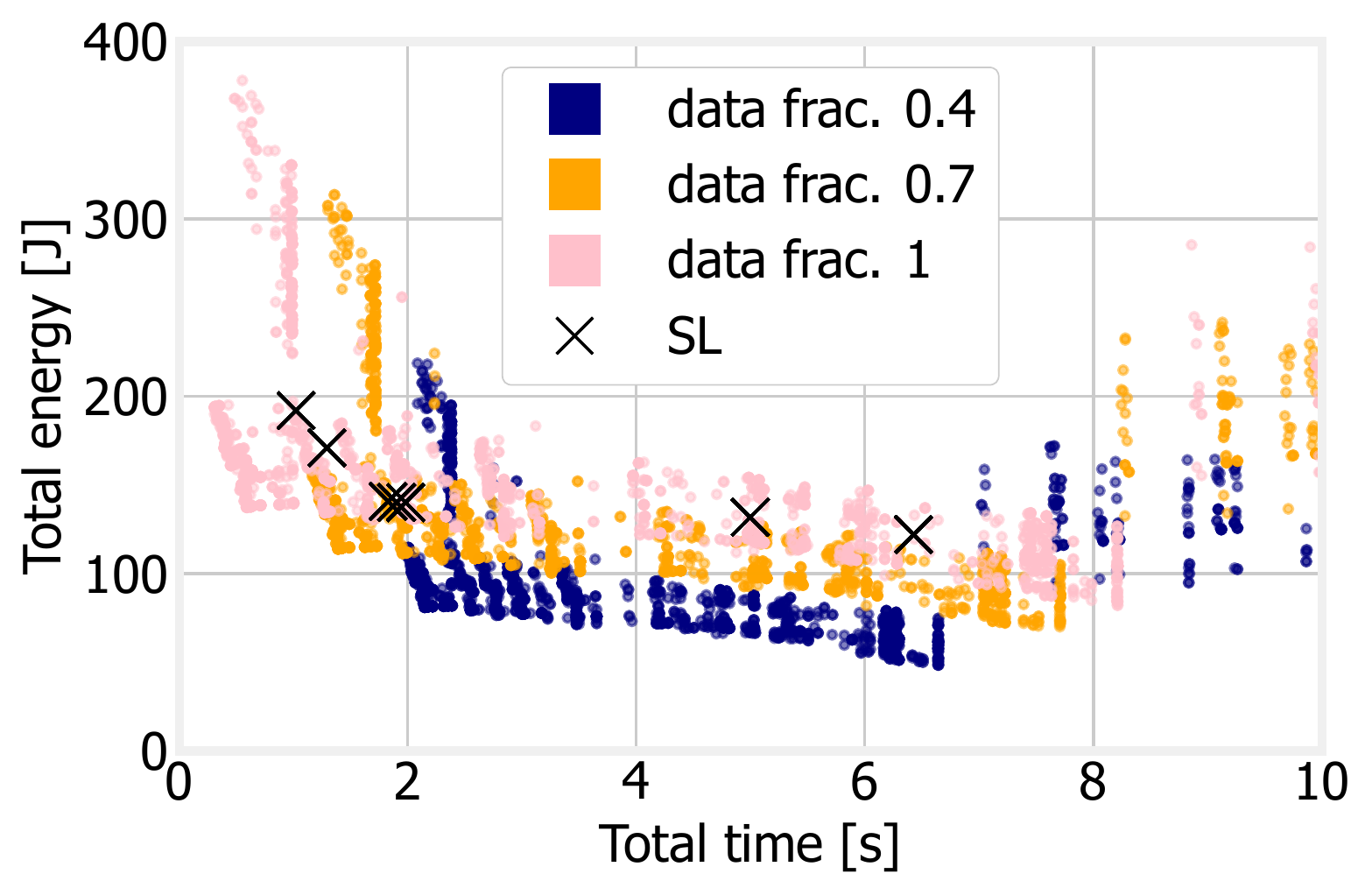}
\caption{
    Small-scale scenario: energy/time trade-offs possible under the RightTrain (dots) and 
    SL (crosses) strategies. Dots of different colors correspond to different fractions of used data.
\label{fig:small-scatter}
} 
\end{figure}

\begin{table}
\caption{
Complexity of the layers of the AlexNet DNN used for our performance evaluation
\label{tab:layers}
} 
\begin{tabularx}{1\columnwidth}{|X|l|r|}
\hline
Layer name & Type & Complexity [MOPs] \\
\hline\hline
conv1 & convolutional & 0.043 \\
\hline
conv2 & convolutional & 6.771 \\
\hline
conv3 & convolutional & 10.145 \\
\hline
conv4 & convolutional & 13.523 \\
\hline
conv5 & convolutional & 9.017 \\
\hline
fc1 & fully-connected & 4.001 \\
\hline
fc2 & fully-connected & 16.027 \\
\hline
fc3 & fully-connected & 0.039 \\
\hline
\end{tabularx}
\vspace{.7cm}
\caption{
Computational capability and power consumption of gold, silver, and bronze servers
\label{tab:nodes}
} 
\begin{tabularx}{1\columnwidth}{|l|X|X|X|X|}
\hline
Class & Real-world example & Capability [TOPS] & Power consumption [W] & Efficiency [W/TOPs] \\
\hline\hline
Gold & NVIDIA Ampere A100~\cite{a100} & 312 & 400 & 1.28\\
\hline
Silver & NVIDIA RTX A4000~\cite{rtxa4000} & 153.4 & 140 & 0.91\\
\hline
Bronze & Apple A14 bionic~\cite{a14} & 11 & 6 & 0.54\\
\hline
\end{tabularx}
\end{table}

\begin{figure*}
\centering
\includegraphics[width=.32\textwidth]{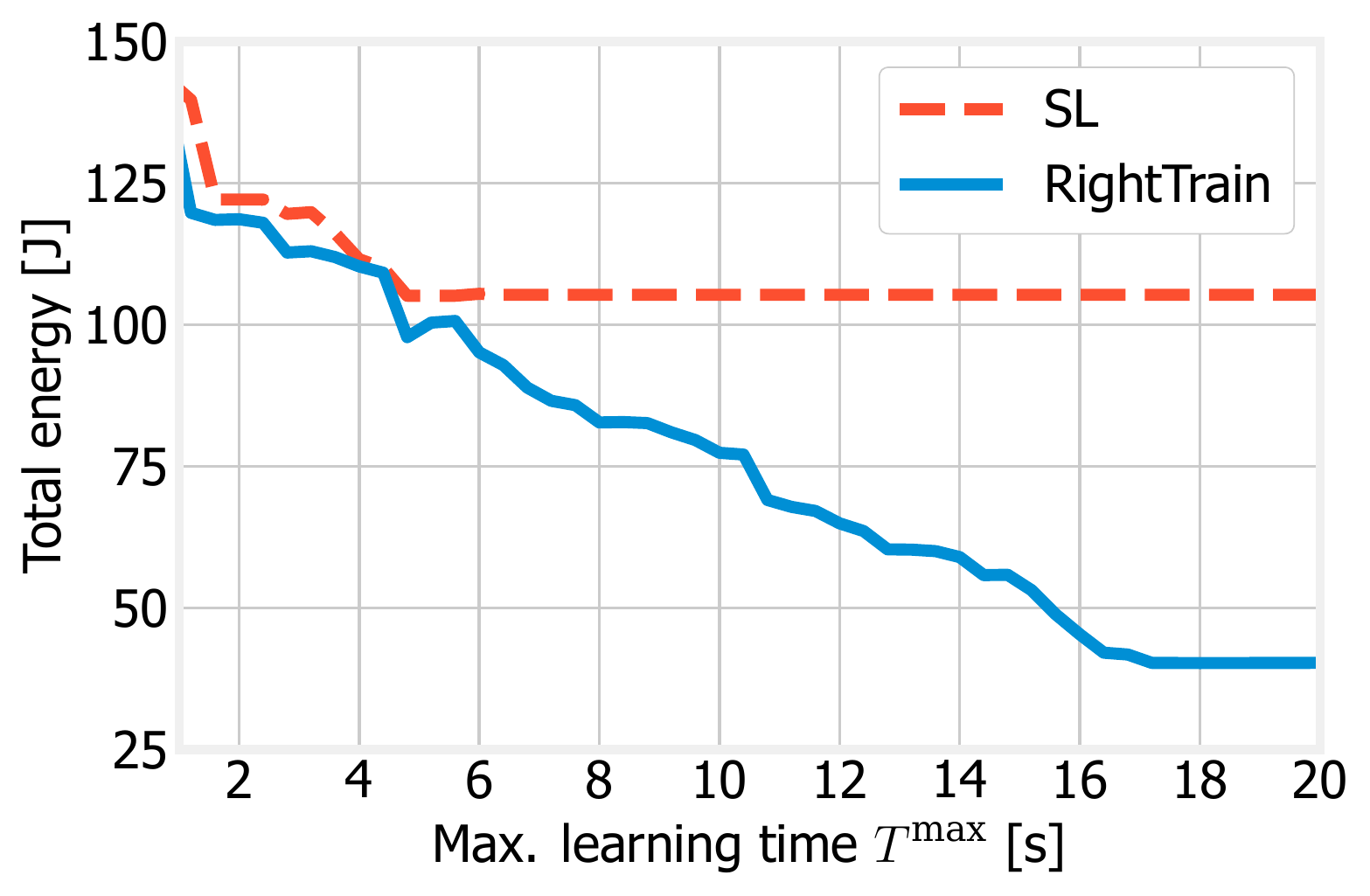}
\includegraphics[width=.32\textwidth]{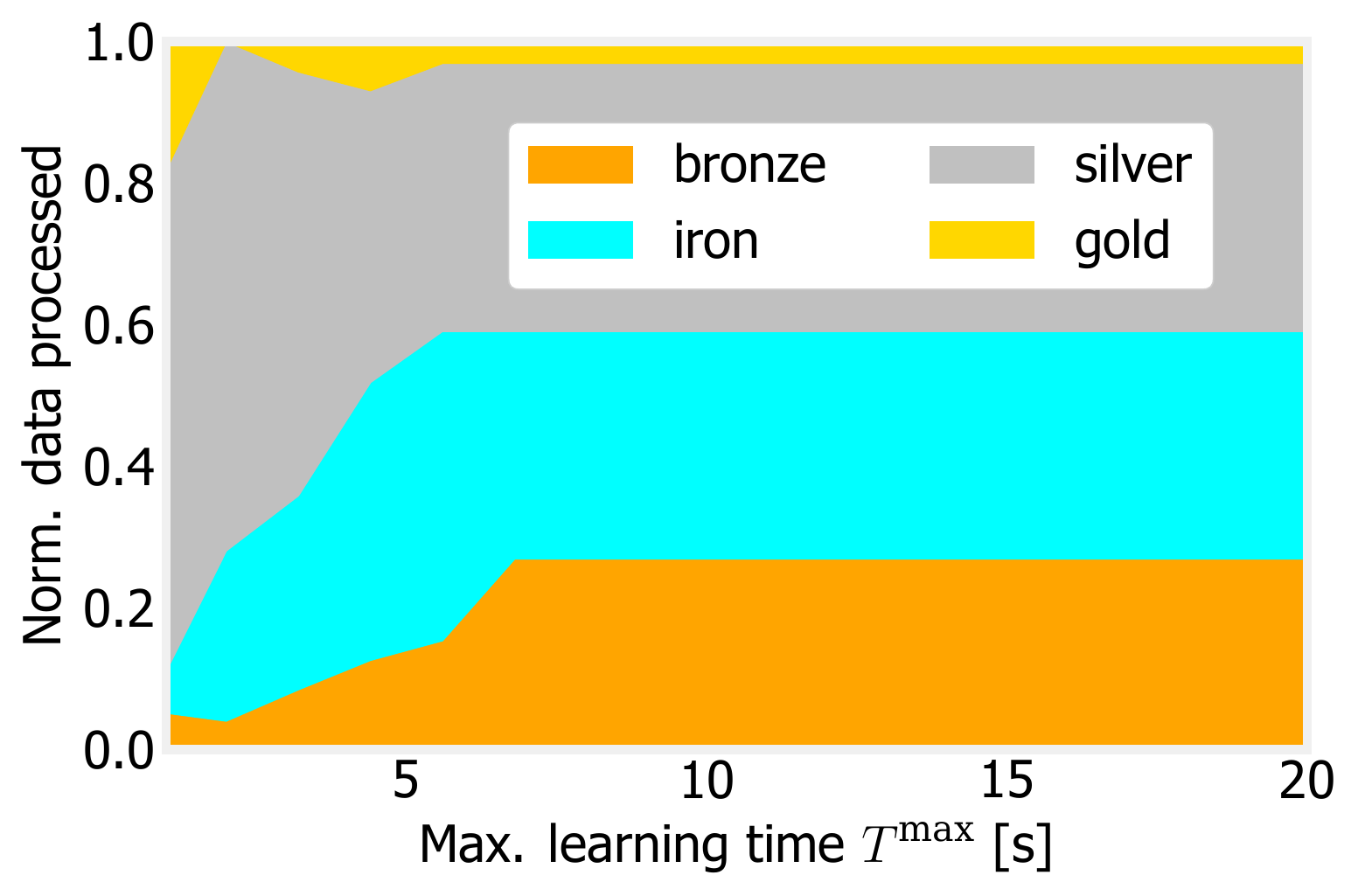}
\includegraphics[width=.32\textwidth]{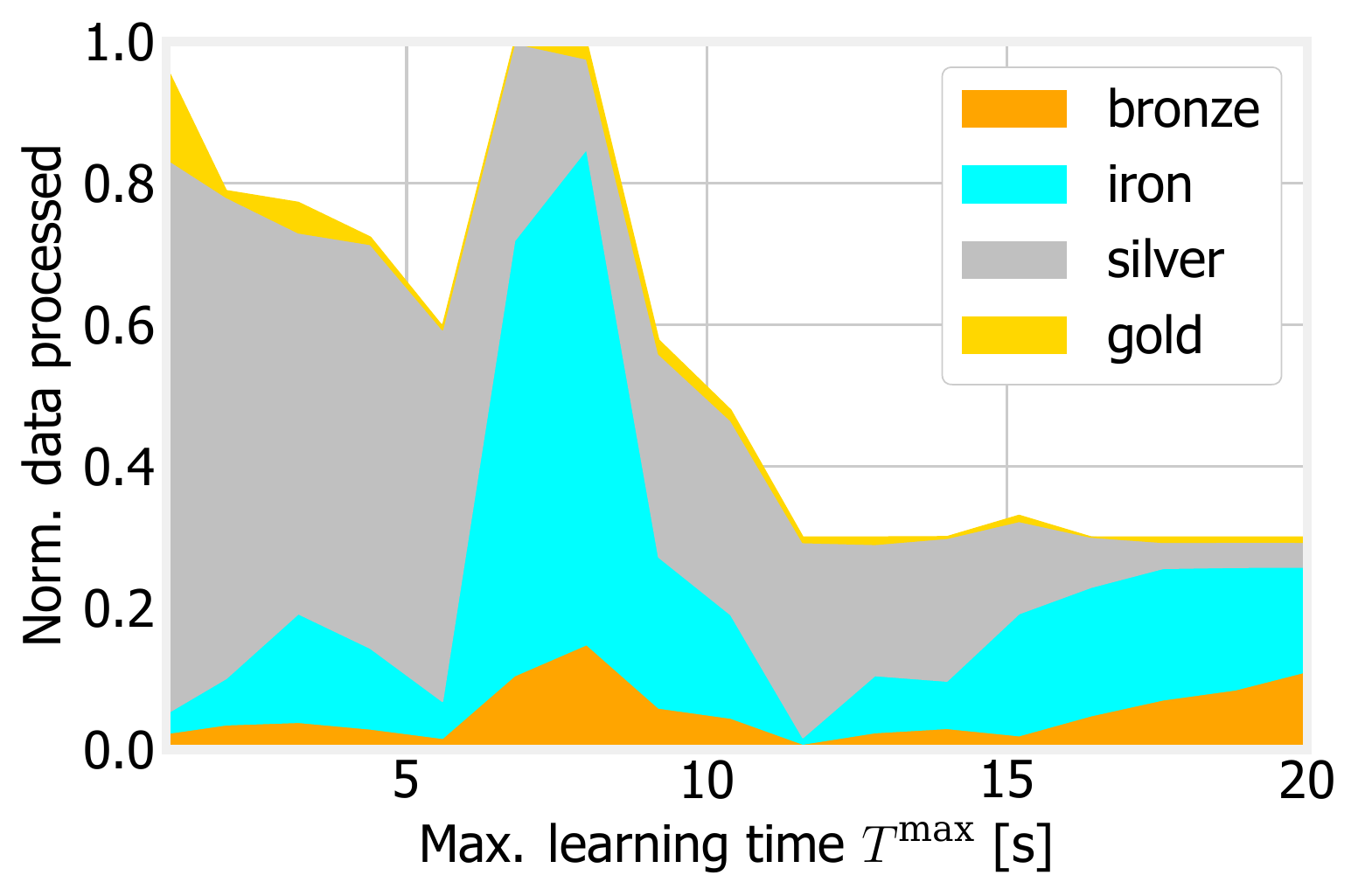}
\caption{
    Large-scale scenario: energy consumed as a function of the maximum learning time~$T^{\max}$ (left), and 
    quantity of data processed at nodes of different classes under the SL (center) and RightTrain (right) strategies.
\label{fig:large-general}
} 
\end{figure*}

{\bf Network scenarios.}
We consider three-tier scenarios like the one exemplified in \Fig{brains}, 
featuring the mobile-edge-cloud continuum and including:
\begin{itemize}
    \item {\em user equipment (UE)}, e.g.,
smart-city devices like cameras and sensors:
    they may produce data (blue cylinders) and/or have computational capabilities;
    \item {\em edge- and cloud-level} datacenters: they contain multiple (virtual) servers.
\end{itemize}
The computational capabilities of UEs are rated {\em bronze}, i.e., very
limited. Edge- and cloud-level servers, instead, come in {\em silver}
and {\em gold} variants, the latter with very large, computational
capabilities.
Importantly, 
as can be seen from \Tab{nodes}, 
lower-capability servers have better efficiency, i.e., need
fewer watts to provide the same number of trillions of operations (TOPs). This suggests that being
able to exploit {\em all} elements of the mobile-edge-cloud continuum,
including using less-powerful devices for moderate loads, is an
important asset for any decision-making strategy.

We begin our performance evaluation from a {\em small-scale scenario},
which allows for a comparison again the optimum (see below); this
includes four data sources and five computation-capable nodes (three
edge servers and two cloud ones). We then move to a {\em large-scale
scenario}, where the number of data sources and nodes grows to 15 and
20, respectively.
Further, in the large-scale scenario we introduce a forth type of
nodes, denoted as {\em iron}, with intermediate features between bronze
and silver ones.

{\bf Benchmark strategies.}
We compare the performance of RightTrain against 
SL~\cite{vepakomma2018split}, owing to its power and
performance~\cite{gao2020end}. Specifically, SL splits the DNN into
three parts, and aims at running one at each of the mobile, edge, and
cloud layers of the network topology; for each layer, the viable server
resulting in the lowest energy consumption is chosen. Since we are
interested in the {\em best} decisions that can be made under the SL
paradigm, we compare all possible splits, and choose the one resulting
in the best value of the objective \Eq{obj}.

Furthermore, as mentioned above, in the small-scale scenario, 
we compare against optimal decisions, obtained by trying all
possible combinations  through brute force.

\subsection{Numerical results}
\label{sec:results}

The most basic aspect in which we are interested is how effective
RightTrain and its counterparts are in pursuing the optimization objective
\Eq{obj}. To this end, \Fig{small-general}(left) shows the energy
consumed as a function of the maximum learning time~$T^{\max}$, for the
small-scale scenario. Consistently with intuition, lower values
of~$T^{\max}$, hence, tighter delay constraints, result in a higher
energy consumption.

As for the relative performance of RightTrain and its alternatives, we can
identify two distinct regions. When $T^{\max}$ is small, hence, delay
constraints are very tight, all strategies perform similarly, with
RightTrain consuming slightly less energy than SL and close to the optimum,
owing to its greater flexibility in making instance-to-node matching
decisions. As $T^{\max}$~increases, we can observe that the energy
associated with SL stops decreasing, while RightTrain is able to track the
optimum and yield a substantially lower energy consumption, 
over 50\% less than SL. 
The reason
for such behavior is shown in \Fig{small-general}(center): SL can
result in learning times that are shorter than~$T^{\max}$, especially
when $T^{\max}$~itself is higher.

One reason for this is shown in \Fig{small-general}(right), portraying
the fraction of data used by each strategy (purple) and the resulting
number of iterations~$K$ (green). We can see that SL (dashed lines)
always uses all available data, which results in a constant (and low)
number of iterations. On the other hand, both RightTrain and the optimum
are able to use less data when the delay restrictions are looser,
achieving a lower energy consumption and, hence, a better efficiency,
in spite of a higher number of iterations.

The second reason is shown in \Fig{small-whereprocess}, depicting how
each strategy utilizes the different parts of the network topology. We
can observe that SL (left plot) tends to use more low-powered mobile
nodes as $T^{\max}$~increases, as one might expect. For RightTrain (center
plot) and the optimum (right plot), the quantity of data to process
decreases as $T^{\max}$~increases, which allows for a greater flexibility in
using all segments of the mobile-edge-cloud continuum, including
high-powered cloud nodes when appropriate. Notice that, under 
RightTrain and (to an even greater extent) the optimal strategy, the curves
in \Fig{small-whereprocess} do not look smooth, e.g., the quantity of
data processed at mobile nodes fluctuates as $T^{\max}$~increases. This
is in contrast with the monotonic evolution in
\Fig{small-whereprocess}(left), and reflects the fact that RightTrain is
better than SL at accounting for the nonlinearities of the system
behavior (e.g., the fixed energy component~$e_\text{f}$) and it adjusts its
decisions accordingly.

\Fig{small-scatter} provides further insights about the greater
flexibility of RightTrain compared to SL. Each marker in the plot
corresponds to a possible solution, with its position along the $x$- and
$y$-axes corresponding, respectively, to its learning time and energy.
Dots represent solutions reachable by RightTrain, with their color
corresponding to the fraction of used data; black crosses represent
solutions reachable by SL. We can immediately see that being able to not
use all data allows RightTrain to explore a larger set of high-quality
trade-offs, often with a smaller energy consumption and longer learning
time. As for SL, all of the solutions it can explore 
can also be reached by RightTrain when all data is used (pink dots).

We now move to the large-scale scenario and plot, in
\Fig{large-general}(left), the energy consumed by the SL and RightTrain
strategies (indeed, owing to the scenario size, computing the optimum is not
feasible). It is possible to observe a similar behavior to that in 
\Fig{small-general}(left), with RightTrain always yielding a smaller power
consumption than SL, and the difference growing as $T^{\max}$~gets
larger. By comparing \Fig{large-general}(left) to 
\Fig{small-general}(left), it is also possible to observe how RightTrain
performs noticeably better than SL even for small values of~$T^{\max}$,
and how the two curves diverge earlier in \Fig{large-general}(left) than
in \Fig{small-general}(left).

The reason for such a different behavior is presented in
\Fig{large-general}(center) and \Fig{large-general}(right), depicting 
how (respectively) SL and RightTrain use the different types of nodes in
the topology. Similarly to \Fig{small-whereprocess}, RightTrain can 
more flexibly -- one would almost say, {\em creatively} -- use available
physical nodes, including ``iron'' ones, thus yielding  a lower energy
consumption than SL. The behavior and performance difference is more
clear here than for the small-scale scenario, due to the wider variety of
existing nodes.

\section{Testbed validation}
\label{sec:testbed}

\begin{figure}
\centering
\includegraphics[width=.9\columnwidth]{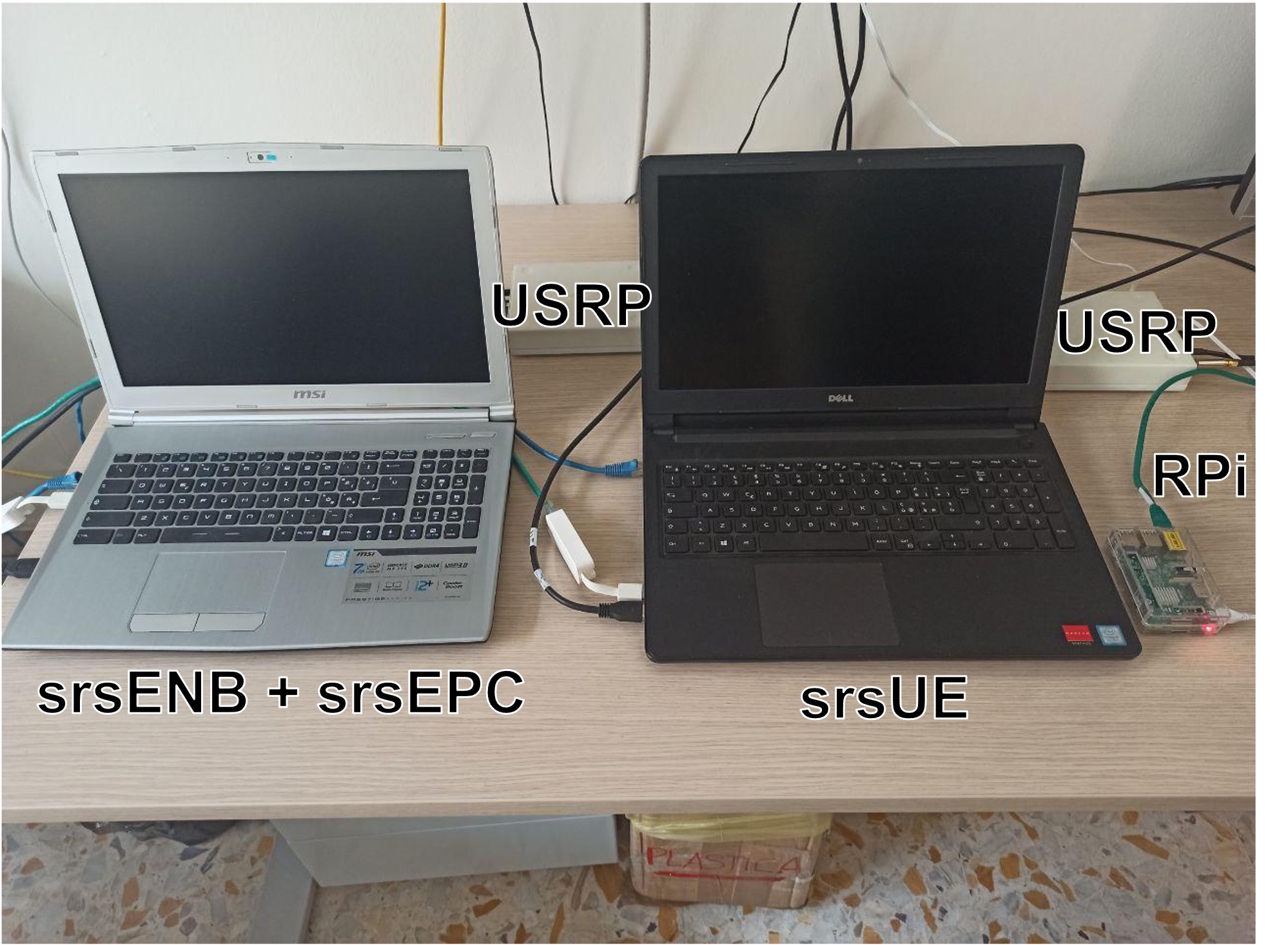}
\caption{
The nodes of the lab test-bed we employ.
    \label{fig:testbed-photo}
} 
\end{figure}

\begin{figure*}[h!]
\centering
\subfigure[\label{fig:bananas2}]{
    \includegraphics[width=.24\textwidth]{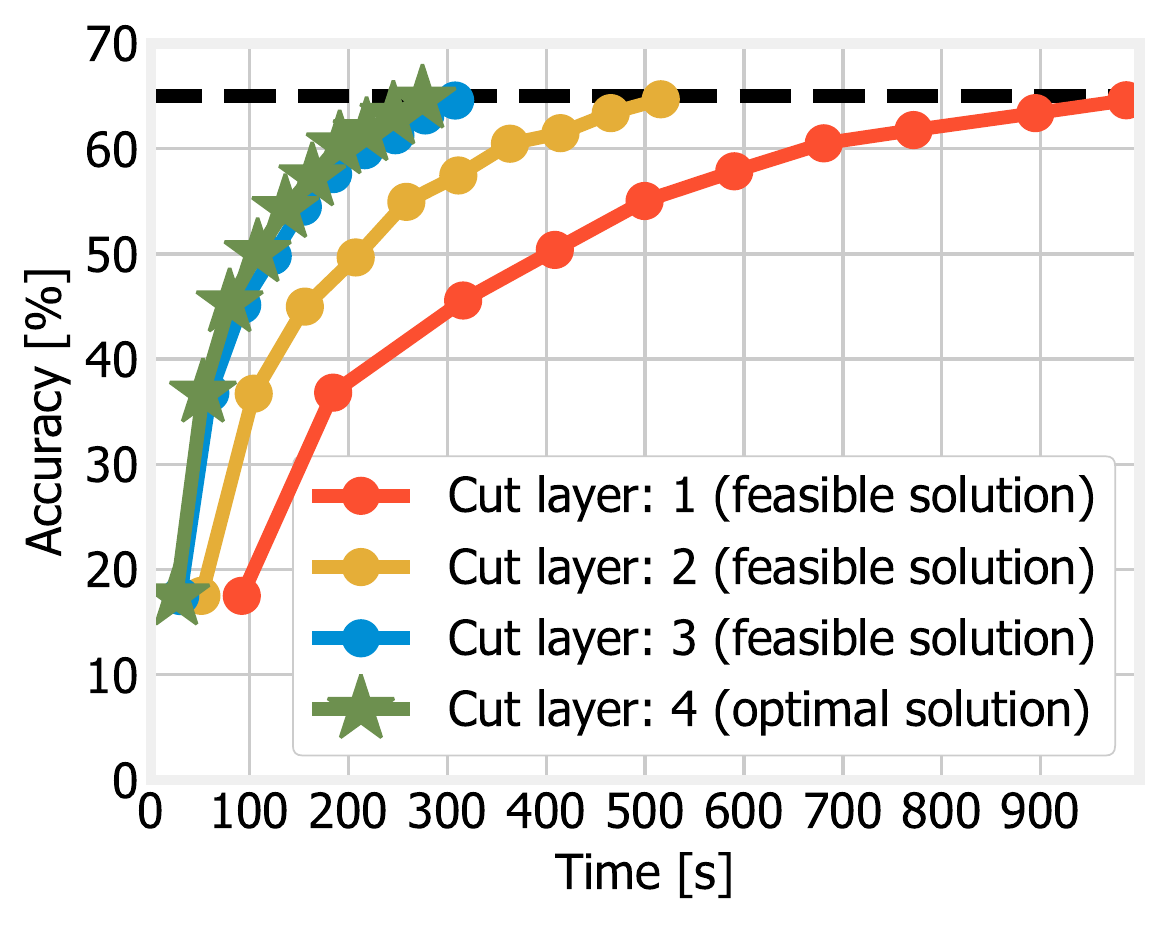}
}\hspace{-2mm}
\subfigure[\label{fig:bars2}]{
    \includegraphics[width=.24\textwidth]{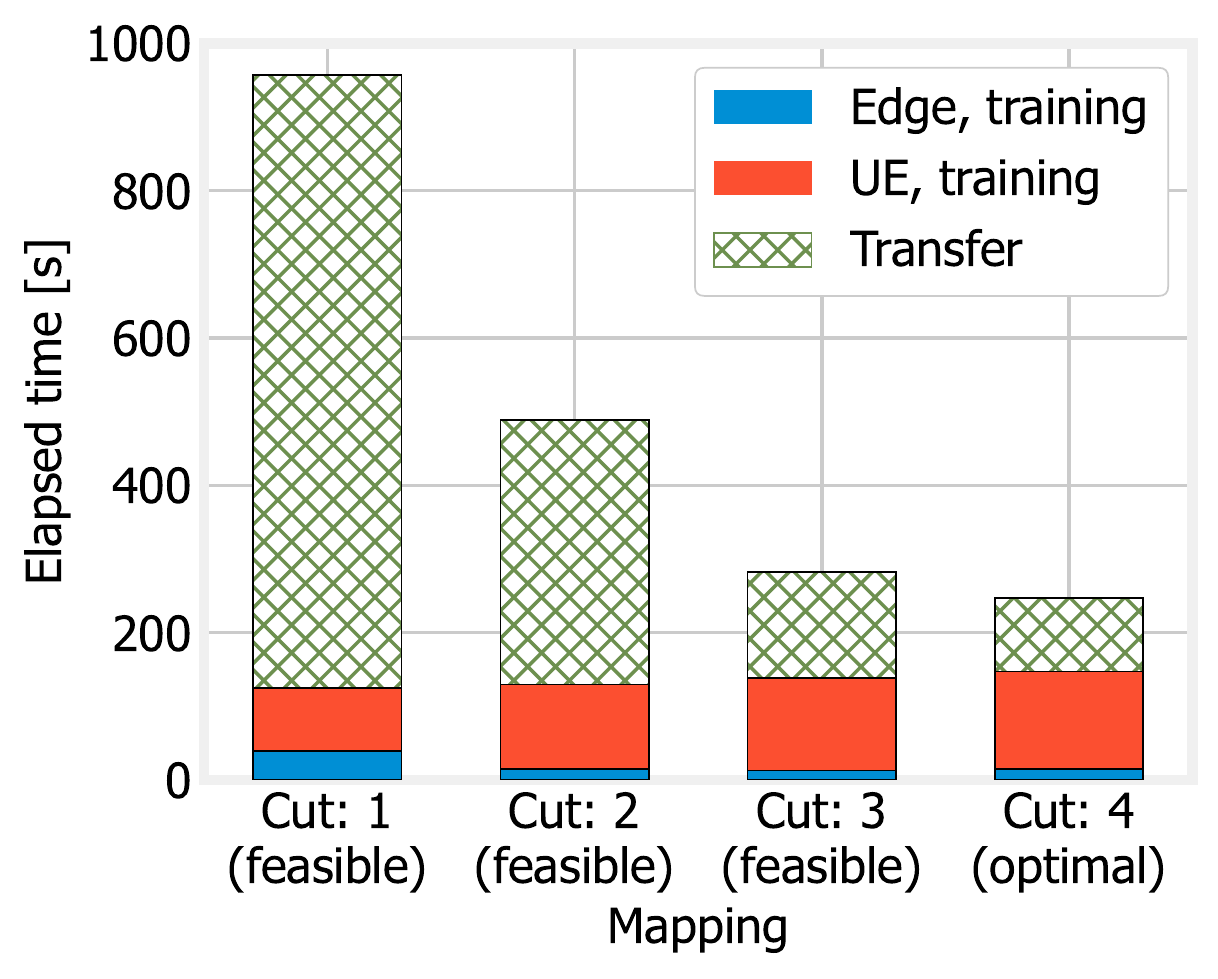}
}\hspace{-2mm}
\subfigure[\label{fig:gantt2slow}]{
    \includegraphics[width=.24\textwidth]{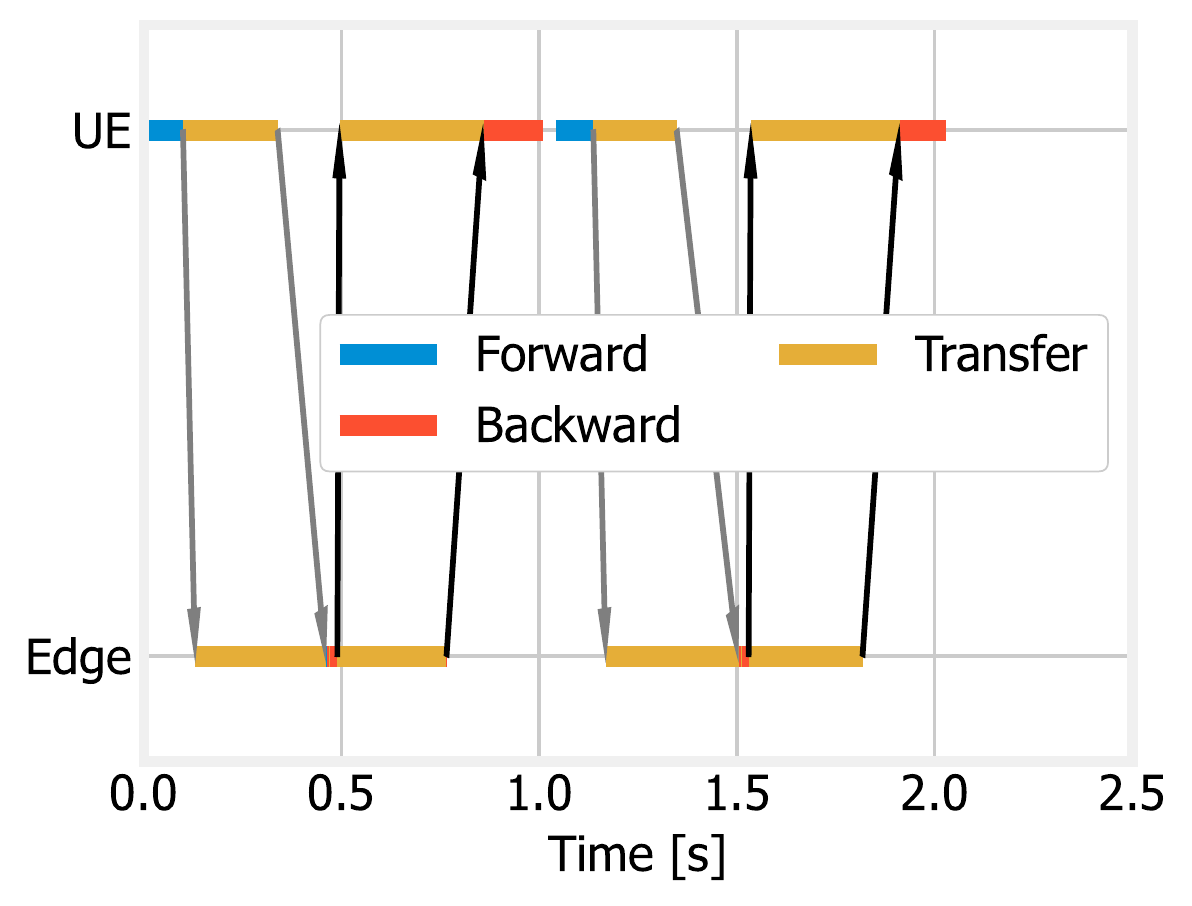}
}\hspace{-2mm}
\subfigure[\label{fig:gantt2fast}]{
    \includegraphics[width=.24\textwidth]{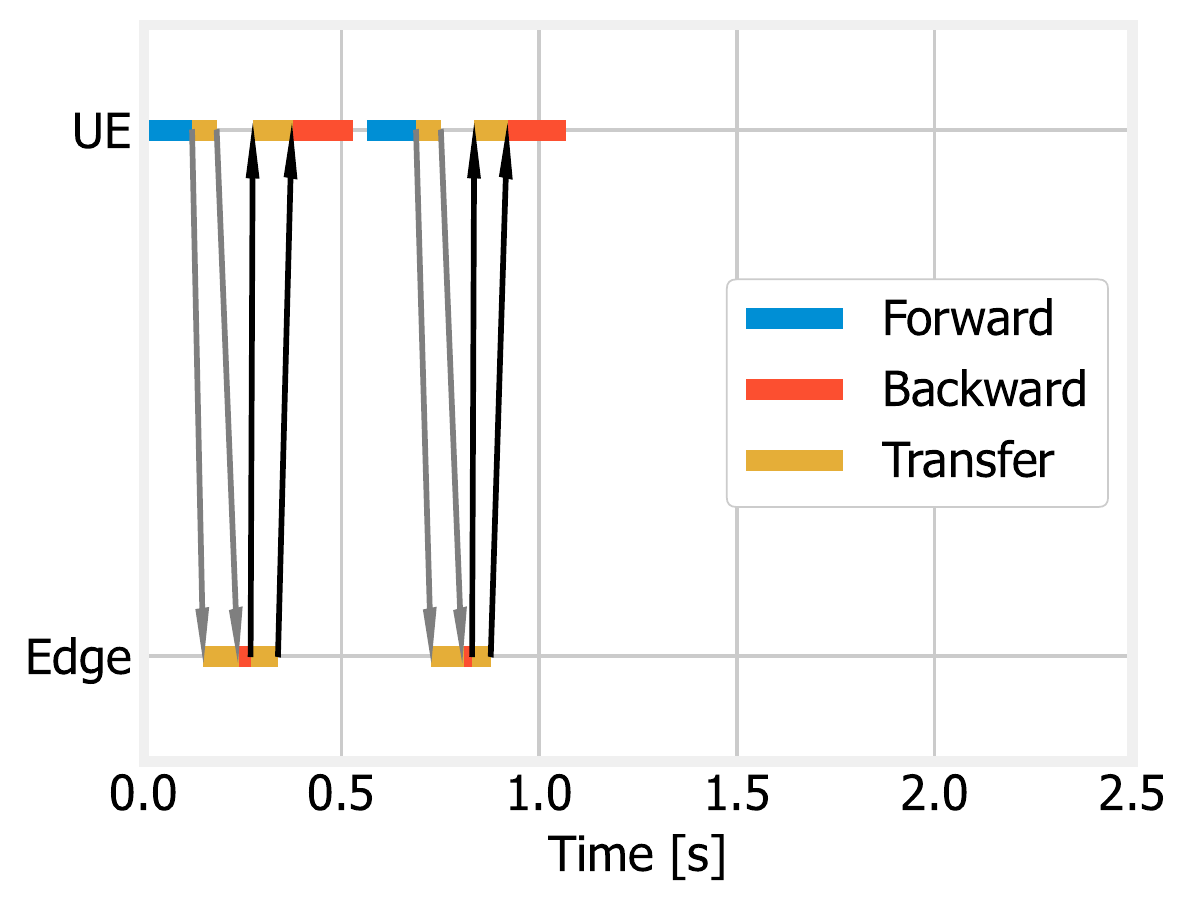}
}
\caption{
Lab test-bed, two-node configuration: accuracy vs. time for different mappings (a); 
total elapsed time for different mappings (b); Gantt chart for the ``cut layer: 2'' (c) and ``cut layer: 4'' (d) mappings.
    \label{fig:testbed2}
} 
\centering
\subfigure[\label{fig:bananas3}]{
    \includegraphics[width=.24\textwidth]{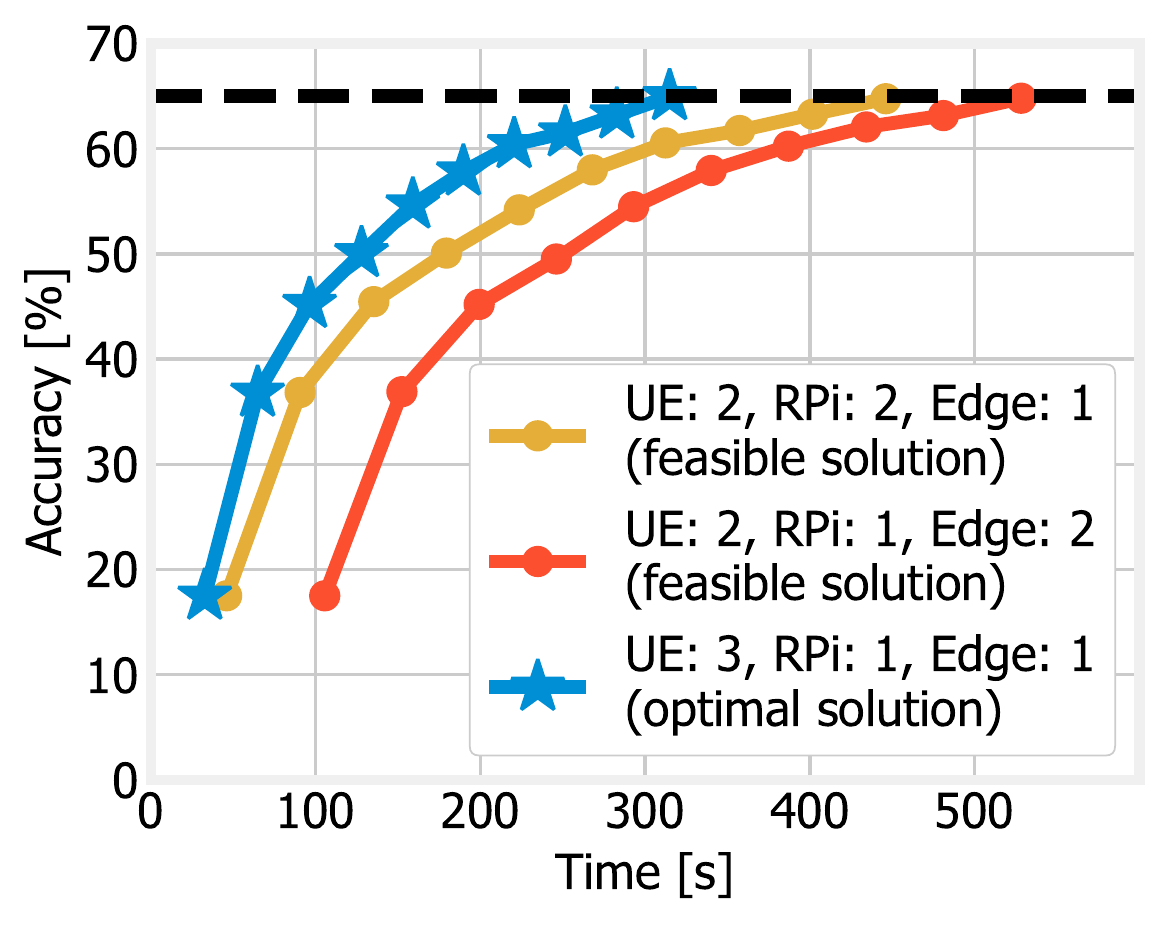}
}\hspace{-2mm}
\subfigure[\label{fig:bars3}]{
    \includegraphics[width=.24\textwidth]{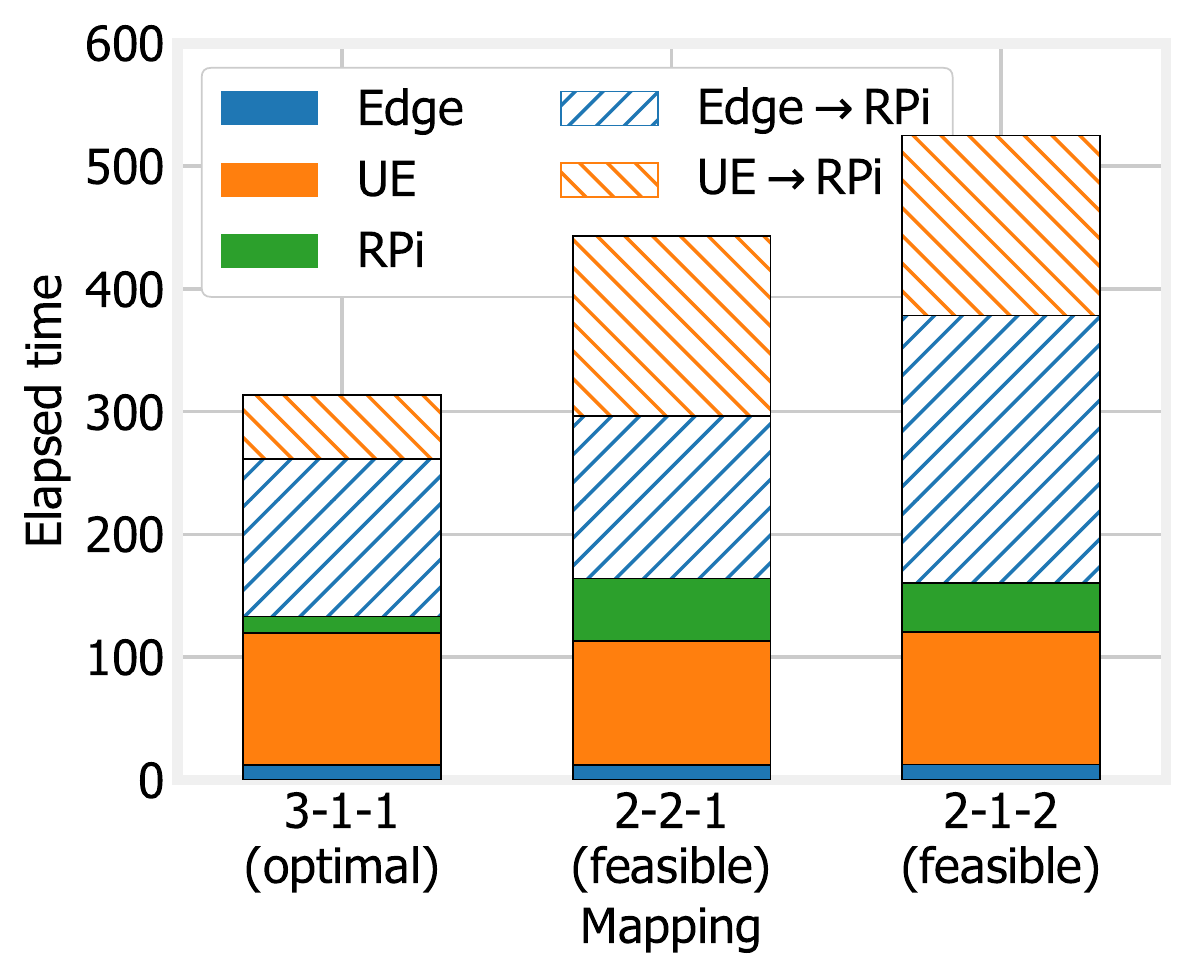}
}\hspace{-2mm}
\subfigure[\label{fig:gantt3slow}]{
    \includegraphics[width=.24\textwidth]{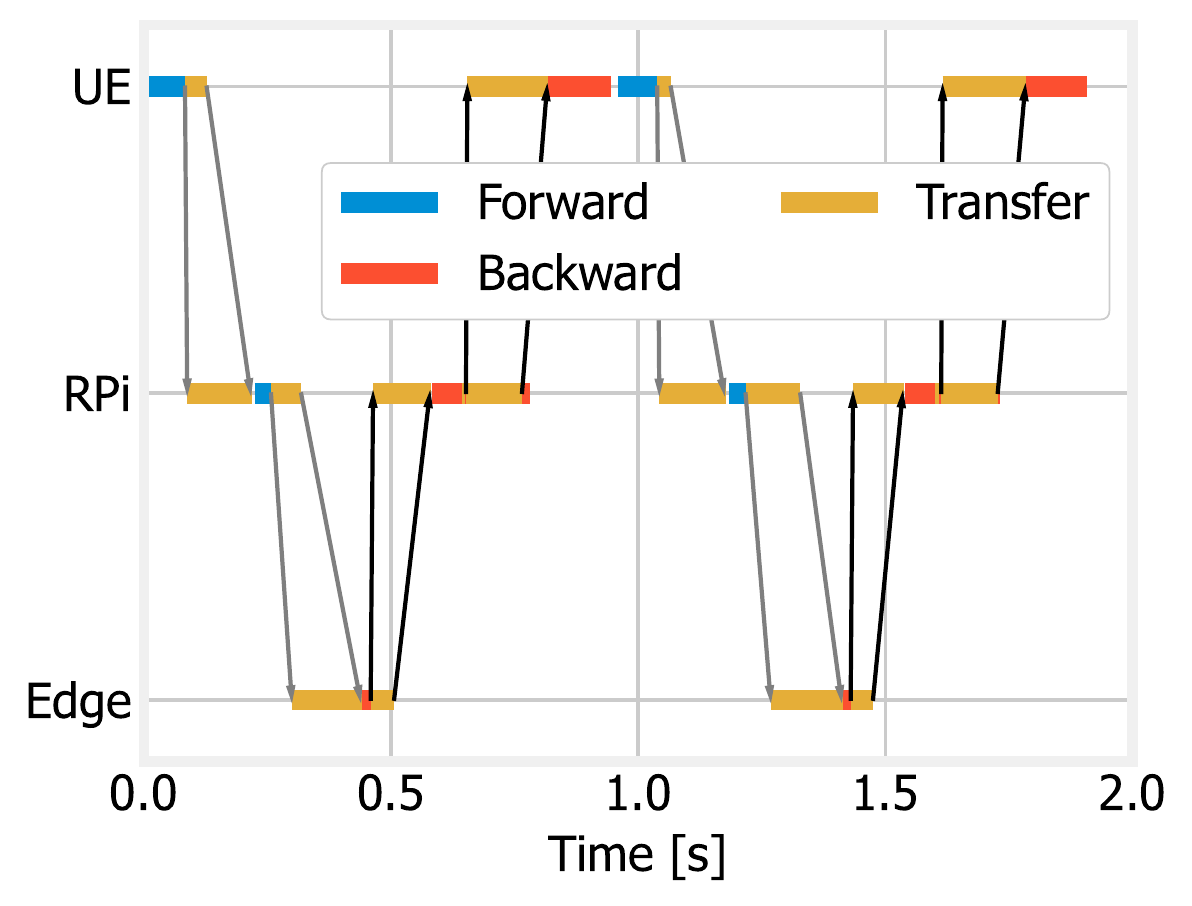}
}\hspace{-2mm}
\subfigure[\label{fig:gantt3fast}]{
    \includegraphics[width=.24\textwidth]{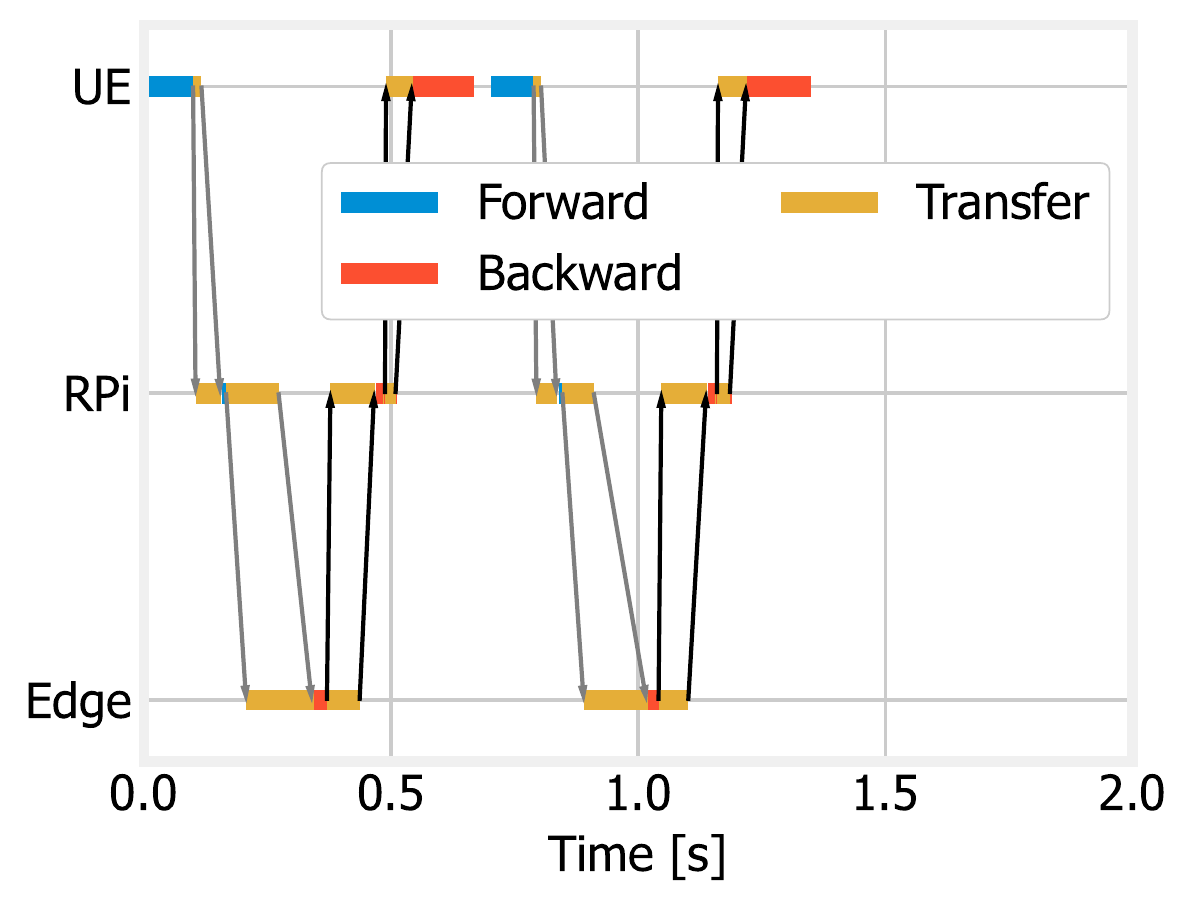}
}
\caption{
Lab test-bed, three-node configuration: accuracy vs. time for different mappings (a); total elapsed time for different mappings (b); Gantt chart for the ``UE: 3, RPi: 1, Edge node: 1'' (c) and ``UE: 2, RPi: 1, Edge node: 2'' (d) mappings.
    \label{fig:testbed3}
} 
\end{figure*}

We now validate our model and approach through a lab test-bed composed of three nodes, depicted in \Fig{testbed-photo}:
\begin{itemize}
    \item a laptop, acting as {\em edge node}, and equipped with an Intel i7-7700HQ CPU and 8~GB of DDR4 RAM;
    \item a second laptop, acting as {\em UE}, equipped with an Intel i7-8550U processor and 16\,GB of DDR4 RAM;
    \item a Raspberry Pi (RPi) 3 Model B, carrying a quad-core 1.2~GHz Broadcom BCM2837 and 1~GB of RAM.
\end{itemize}
Laptops run the Ubuntu 18.04 operating system, while the RPi runs Ubuntu Server 20.04. UE and edge node are 
connected through a 3GPP LTE virtualized Radio Access Network (vRAN), 
leveraging Ettus Universal Software Radio Peripheral (USRP) B210 boards. 
The vRAN is based on the srsRAN~\cite{srsRAN} open-source LTE stack implementation, which is compliant with LTE Release 9. 
The RPi is connected to the UE through Wi-Fi, with the latter acting as an access point.

As the learning activity to perform, we consider an image classification task over the CIFAR-10~\cite{krizhevsky2009learning} dataset using the Lenet DNN~\cite{lenet}, composed of two convolutional layers followed by three linear ones. We study the performance and behavior of the possible (i.e., feasible) {\em mappings} between layers and physical nodes, under two test-bed configurations:
\begin{itemize}
    \item a two-node configuration, where only the Edge node and EU are included and different mapping decisions also 
    imply {\em cutting} the DNN after a different number of layers;
    \item a three-node configuration, where the RPi is also used, hence, mapping decisions can be more complex.
\end{itemize}
In all cases, the target accuracy is set to 65\%, and the maximum learning time is 1,000~s.

\Fig{testbed2} reports the results for the two-node configuration. 
From \Fig{bananas2}, we can observe that 10~epochs are always sufficient to reach the target accuracy; 
however, the {\em time} needed to perform such epochs changes significantly; specifically, the later we ``cut'' the network, the shorter the learning time. 
The reason of this behavior is highlighted in \Fig{bars2}. Interestingly, the total computing time (i.e., considering both the Edge node and the UE) remains roughly constant 
(since the Edge node and UE laptops have similar performance). On the contrary, 
the amount of data to be transmitted, hence, the data transfer time, 
decreases substantially (up to 73\%) with higher values of the cut layer, i.e., if we cut the network 
after a larger number of layers. 
This is consistent with the fact that later layers of the DNN (i.e., farther from input data) exchange less data, hence, ``cutting'' the DNN at such layers reduces the quantity of information to exchange between nodes.

\Fig{gantt2slow} and \Fig{gantt2fast} cast additional light on this phenomenon, by depicting how the two nodes alternate performing computations and exchanging data in the first two batches of a typical iteration. For each batch, each node performs the forward step and transmits the output of its own last layer to the following node. 
Then, when the Edge node terminates the forward stage of the last layer, it computes the loss and starts the backward procedure, computing the gradients and sending them back. The transmissions of the output and gradients are indicated respectively by the gray and black arrows in the plots. Blue and red bars therein correspond to forward and backward passes, with the latter always taking roughly twice as much as the former. It is easy to notice that, while the forward and backward passes take roughly the same time, ``cutting'' the DNN at layer~2 (\Fig{gantt2slow}) instead of layer~4 (\Fig{gantt2fast}) results in a much larger quantity of data to transmit, hence, longer total training times.

\Fig{testbed3} confirms the findings above, in spite of the fact that the DNN layers can be spread across three nodes, hence, mappings are more complex. As in \Fig{bananas2}, in \Fig{bananas3} the accuracy reached at each epoch does not change, but the time such epochs take does depend upon the mapping. Importantly, such a time is deeply influenced, as shown in \Fig{bars3}, by the transfer times between the nodes.

\section{Conclusion and Future Work}
\label{sec:conclusion}

We have addressed  distributed training of DNNs in the
mobile-edge-cloud continuum, and identified the challenge of making 
joint, energy-efficient decisions on such diverse aspects as (i) selecting
the data to be used, (ii) choosing the distributed DNN structure, and (iii)
matching DNN layers with the physical nodes to run them. We have
presented a solution concept, centered around the RightTrain algorithm,
making all necessary decisions  in polynomial time and within
$2(1+\epsilon)$~from the optimum, with the objective of minimizing the
total energy consumption. Our performance evaluation  shows that 
RightTrain  closely
matches the optimum and reduces  the energy consumption of a learning task by over 50\% with respect to 
the state of the art. Furthermore, we have validated our approach by implementing it in a lab test-bed. 

Future work will include testing RightTrain against hybrid
supervised/unsupervised learning tasks, so as to assess its
effectiveness and performance when faced with even larger DNNs and more
complex learning techniques. Also, it would interesting to investigate the synergy between RightTrain 
and model-compression techniques like pruning and knowledge distillation. 

\bibliographystyle{IEEEtran}
\bibliography{refs}

\end{document}